\documentclass[jgrga]{agu2001}
\usepackage{graphicx}
\usepackage{subfigure}
\usepackage{verbatim}
\usepackage{textcomp}
\usepackage{epstopdf}

\RequirePackage{bm,amsmath,xspace}
\allowdisplaybreaks
\newcounter{regequation}
\newcounter{mycounter}
\newcommand{\newcm}[2]{\newcommand*{#1}{\ensuremath{#2}}}

\newcommand{\newc}{\newcommand*}

\providecommand{\what}{\widehat}
\providecommand{\wtilde}{\widetilde}

\renewenvironment{quote}
 {\begin{list}{}
 {
 \setlength{\rightmargin}{\leftmargin}}
 \item[]}%
  {\end{list}}

  {\setcounter{equation}{\value{regequation}}
    \end{minipage}\end{shaded}\end{minipage}\end{center}\end{table} }

  {\setcounter{equation}{\value{regequation}}
   \end{minipage}\end{shaded}\end{minipage}\end{center}\end{table} }

 {\setcounter{equation}{\value{regequation}}
   \end{boxedminipage}\end{center}\end{table} }

    {\setcounter{equation}{\value{regequation}}
      \end{boxedminipage}\end{center}\end{table} }

{\vskip 2pt \begin{compactitem}[#1]\setlength{\itemsep}{2pt}}%
{\end{compactitem}\vskip 2pt}

{\vskip 2pt \begin{compactenum}[#1]\setlength{\itemsep}{2pt}}%
{\end{compactenum}\vskip 2pt}

{\begin{enumerate}[#1]}{\end{enumerate}}

{\begin{enumerate}[#1]}{\end{itemize}}

  {\newpage \end{list} \end{small} \end{sffamily}  }
   {\newpage \end{list} \end{small} \end{sffamily}}
  
{\end{list}\end{sffamily}\end{small}}

{\end{list}\end{sffamily}\end{small}}

 {\end{list}}

 {\end{list}}

 {\end{list}}

 {\end{list}}

\newcommand{\eqnab}%
    {\renewcommand{\theequation}{\arabic{chapter}.\arabic{equation}a,b}}
\newcommand{\eqnabc}%
    {\renewcommand{\theequation}{\arabic{chapter}.\arabic{equation}a,b,c}}

\newcommand{\aleqnab}%
	{\renewcommand{\theequation}{\theparentquation\arabic{equation}a,b}}

\newcommand{\itsf}{\itshape\sffamily}

\def\gobble #1{}   

\newc{\x}{$x$\xspace} \newc{\y}{$y$\xspace}  \newc{\z}{$z$\xspace}

\newc{\afterpag}{\afterpage}

\newc{\probref}[1]{\thechapter.\ref{#1}}

\newc{\letrn}[2]{\lettrine[lines=3,lhang=0]{{\color{gray}#1}}
               {\MakeUppercase{\footnotesize\sffamily #2}}}
\newc{\figref}[1]{Fig.\ \ref{#1}}
\newc{\Figref}[1]{Fig.\ \ref{#1}}
\newc{\secref}[1]{section \ref{#1}}

\newc{\dbint}{\iint}

\newc{\curlz}{\mathrm{curl}_z}

\newcommand{\dg}{\text{\textdegree}\xspace}

\bmdefine{\omb}{\omega}
\bmdefine{\Omb}{\varOmega}

\bmdefine{\taub}{\tau}
\bmdefine{\deltab}{\delta}
\newc{\ttaub}{\wtilde{\taub}}
\newc{\ttau}{\wtilde\tau}

\newc{\s}{\,\text{s}\xspace} \newc{\m}{\,\text{m}\xspace}
\newc{\cm}{\,\text{cm}\xspace}
\newc{\km}{\,\text{km}\xspace}
\newc{\kg}{\,\text{kg}\xspace} \newcm{\ps}{\, \text{s}^{-1}\xspace}
\newc{\Kv}{\,\text{K}\xspace}
\newc{\Sv}{\,\text{Sv}\xspace}
\newc{\rms}{\text{rms}}
\newc{\qg}{quasi-geostrophic\xspace}
\newc{\pg}{planetary geostrophic\xspace}
\newc{\BV}{Brunt-V\"ais\"al\"a\xspace}
\newc{\crn}{\nonumber  \\}
\newc{\qqquad}{\qquad \qquad}
\newc{\cdel}{\cdot \nabla}

\newc{\Ro}{\ensuremath{\mathit{Ro}}\xspace}

\newc{\Ek}{\ensuremath{\mathit{Ek}}\xspace}
\newc{\Ra}{\ensuremath{\mathit{Ra}}\xspace}
\newc{\Fr}{\ensuremath{\mathit{Fr}}\xspace}
\newc{\Bu}{\ensuremath{\mathit{Bu}}\xspace}
\newc{\Ri}{\ensuremath{\mathit{Ri}}\xspace}
\newc{\Rot}{\ensuremath{\mathit{Ro_\mathrm{T}}}\xspace}

\newc{\bpp}[2]{\bfrac{\partial #1}{\partial #2}}
\newc{\bppp}[2]{({\partial #1}/{\partial #2})}
\newc{\bfrac}[2]{\left( \frac{#1}{#2} \right) }
\newc{\dbfrac}[2]{\left( \dfrac{#1}{#2} \right) }
\newc{\bfracs}[3]{\left(\frac{#1}{#2}\right)_{\! \! #3}}
\newc{\bfracsub}[3]{\left(\frac{#1}{#2}\right)_{\! \! #3}}
\newc{\bfracsup}[3]{\left(\frac{#1}{#2}\right)^{\! \! #3}}
\newc{\bpps}[3]{\left(\frac{\partial #1}{\partial #2}\right)_{\! \! #3}}
\newc{\bppsub}[3]{\left(\frac{\partial #1}{\partial #2}\right)_{\! \! #3}}
\newc{\bppsup}[3]{\left(\frac{\partial #1}{\partial #2}\right)^{\! \! #3}}
\newc{\bdds}[3]{\left(\frac{\d #1}{\d #2}\right)_{\! \! #3}}

\newc{\Eta}{\mathcal{E}}
\newc{\Zeta}{\mathcal{Z}}

\newc{\K}{\mathcal{K}}

\newc{\psisp}[1]{\psi^{(#1)}}

\newc{\psibar}{\overline \psi}
\newc{\Psibar}{\overline \Psi}
\newc{\phibar}{\overline \phi}
\newc{\thetabar}{\overline \theta}
\newc{\varthetabar}{\overline \vartheta}
\newc{\varphibar}{\overline \varphi}
\renewcommand{\hbar}{\overline h}

\newc{\qbar}{\overline q}
\newc{\Qbar}{\overline Q}
\newc{\ubar}{\overline u}
\newc{\vbar}{\overline v}

\newc{\ubbar}{\overline \ub}
\newc{\vbbar}{\overline \vb}
\newc{\bbar}{\overline b}
\newc{\pbar}{\overline p}
\newc{\zbar}{\overline z}
\newc{\rhobar}{\overline \rho}
\newc{\Fbar}{\overline F}
\newc{\Jbar}{\overline J}
\newc{\Sbar}{\overline S}
\newc{\bybar}{\overline \by}

\newc{\zetabar}{\overline \zeta}
\newc{\etabar}{\overline \eta}
\newc{\betahat}{\ensuremath{\what \beta}}
\newc{\kbeta}{k_\beta}
\newc{\that}{{\what t}}
\newc{\khat}{\what k}
\newcommand{\phat}{\what p}
\newc{\zetahat}{\ensuremath{\what \zeta}}
\newc{\deltahat}{\what \delta}
\newc{\etahat}{\what \eta}
\newc{\ubhat}{\what\ub}
\newc{\vbhat}{\what\vb}
\newc{\xhat}{{\what x}}
\newc{\xbhat}{{\what \xb}}
\newc{\yhat}{{\what y}}
\newc{\zhat}{{\what z}}
\newc{\uhat}{\what u}
\newc{\vhat}{\what v}

\newc{\bhat}{\what b}
\newc{\fbhat}{\what{\bm{f}}}
\newc{\phihat}{\what \phi}
\newc{\byhat}{\what \by}
\newc{\hhat}{\ensuremath{\what h}}
\newc{\fhat}{\ensuremath{\what f}}
\newc{\psihat}{\what \psi}
\newc{\Psihat}{\what \Psi}
\newc{\thetahat}{\what \theta}
\newc{\Nhat}{\what N}
\newc{\Nbar}{\overline N}
\newc{\omegahat}{\what \omega}
\newc{\qhat}{\what q}

\newc{\tauhat}{\what \tau}

\newc{\ug}{u_g}
\newc{\vg}{\ensuremath{v_g}\xspace}
\newc{\Lr}{\Ld}
\newcommand{\Ld}{\ensuremath{L_{d}}\xspace}

\newc{\lr}{\ensuremath{l_r}}
\newc{\lrr}{\ensuremath{\lambda_r}}
\newc{\vrt}{\vartheta}
\newc{\cross}{\bm{\times}}
\newc{\half}{\frac 12}
\newc{\third}{\frac 13}
\newc{\quarter}{\frac 14}

\newc{\dt}{\, \mathrm{d}t}
\newc{\dx}{\, \mathrm{d}x}
\newc{\dy}{\, \mathrm{d}y}
\newc{\dz}{\, \mathrm{d}z}
\newc{\dk}{\, \mathrm{d}k}
\newc{\dr}{\, \mathrm{d}r}
\newc{\dkb}{\, \mathrm{d}\bm{k}}
\newc{\dxb}{\, \mathrm{d}\xb}
\newc{\dvrt}{\, \mathrm{d}\vrt}
\newc{\dpr}{\, \mathrm{d}p}

\newc{\dAb}{\, \mathrm{d} \bm{A}}
\newc{\Dth}{{\D_h \over \D t}}
\newc{\Dto}{{\D_0 \over \D \that}}
\newc{\Fb}{\bm{F} }
\bmdefine\Fb{F}

\newc\Ab{\ensuremath{\bm{A}}}
\newc\Cb{\ensuremath{\bm{C}}}
\newc\Jb{\ensuremath{\bm{J}}}
\newc\Bb{\ensuremath{\bm{B}}}
\newc\xb{{\bm{x}}}
\bmdefine{\pb}{p}
\bmdefine{\qb}{q}
\bmdefine{\ab}{a}
\bmdefine{\Xb}{X}
\bmdefine{\Yb}{Y}
\bmdefine{\rb}{r}

\bmdefine{\sb}{s}
\bmdefine{\Db}{D}

\def\ib{{\bf i}}
\def\jb{{\bf j} }
\newcommand{\kb}{{\bf k}}
\bmdefine{\kbi}{k}
\bmdefine{\Vb}{V}
\bmdefine{\Wb}{W}

\newc{\fb}{{\ensuremath{\bm {f}}}}
\newc{\fbo}{{\ensuremath{\bm {f}_0}}}
\def\vb{{\ensuremath{\bm {v}}}\xspace}
\def\ub{{\ensuremath{\bm {u}}}\xspace}

\bmdefine{\Ub}{U}
\bmdefine{\cb}{c}
\bmdefine{\psib}{\psi}
\bmdefine{\phib}{\phi}

\newcommand{\by}{\ensuremath{b}}

\newc{\Tb}{{\bm T} }
\newc{\lhs}{left-hands side\xspace}
\newc{\fo}{\ensuremath{f_0}\xspace}

\newc{\MD}[1]{{\partial {#1} \over \partial t} + (\vb \cdot \del) #1}
 
\newcommand{\pp}[3][]{{\partial^{#1} #2 \over \partial #3^{#1}}}
\newcommand{\ppp}[3][]{{\partial^{#1} #2 / \partial #3^{#1}}}

\newc{\ppa}[1]{{\partial \over \partial #1}}
\newc{\pt}{{\partial \over \partial t}}
\newc{\del}{\nabla }
\newc{\grad}{\nabla}
\newc{\delh}{\nabla_h}
\newc{\cp}{c_{p}}
\newc{\cv}{c_{v}}

\renewcommand{\d}{\mathrm{d}}

\newcommand{\D}{\mathrm{D}}

\newcommand{\textmonth}{\ifcase\month \or January \or February\or
March\or April\or May \or June \or July\or August\or September\or October
\or November\or December\fi\xspace}

\DeclareMathAlphabet{\mathsfsl}{\encodingdefault}{hlst}{b}{sl}   

\newc{\putabcd}{\put(-9.2,7){\small \textit{\textsf{(a)}}}
\put(-4.5,7){\small \textit{\textsf{(b)}}}
\put(-9.2,3.3){\small \textit{\textsf{(c)}}}
\put(-4.5,3.3){\small \textit{\textsf{(d)}}}
}

\newc{\putab}{\put(-8.9,3.3){\small \textit{\textsf{(a)}}}
\put(-4.3,3.3){\small \textit{\textsf{(b)}}}
}

\bmdefine{\Ebc}{\mathcal E}

\newcommand{\bfig}{\begin{figure}}
\newcommand{\efig}{\end{figure}}
\newcommand{\bcen}{\begin{center}}
\newcommand{\ecen}{\end{center}}
\newcommand{\beq}{\begin{equation}}
\newcommand{\eeq}{\end{equation}}

\newcommand{\bsub}{\begin{subequations}}
\newcommand{\esub}{\end{subequations}}

\renewcommand{\Omega}{\varOmega}
\renewcommand{\Gamma}{\varGamma}
\renewcommand{\Pi}{\varPi}

\usepackage{microtype}
\numberwithin{equation}{section}

\newcommand{\That}{\widehat T}
\newcommand{\Phihat}{\widehat \Phi}

\authorrunninghead{J. L. MITCHELL, G. K. VALLIS}

\titlerunninghead{TRANSITION TO SUPERROTATION}

\authoraddr{Jonathan L. Mitchell,
Department of Earth and Space Sciences,
University of California, Los Angeles,
595 Charles Young Drive East,
Box 951567,
Los Angeles, CA 90095-1567 \newline
(jonmitch@ucla.edu)}

\begin{document}

\title{The Transition to Superrotation in Terrestrial Atmospheres}

\author{Jonathan L. Mitchell}
\affil{Earth and Space Sciences, Atmospheric and Oceanic Sciences and IGPP, UCLA, Los Angeles, CA, USA}
\affil{Institute for Advanced Study, Princeton, New Jersey, USA}
\author{Geoffrey K. Vallis}
\affil{Geophysical Fluid Dynamics Laboratory, Princeton University,
Princeton, New Jersey, USA}

\begin{abstract}
We show that by changing a single non-dimensional number, the thermal Rossby number, global atmospheric simulations with only axisymmetric forcing pass from an Earth-like atmosphere to a superrotating atmosphere that more resembles the atmospheres of Venus or Titan. The transition to superrotation occurs under conditions in which equatorward-propagating Rossby waves generated by baroclinic instability at intermediate and high latitudes are suppressed, which will occur when the deformation radius exceeds the planetary radius.  At large thermal Rossby numbers following an initial, nearly axisymmetric phase, a global baroclinic wave of zonal wavenumber one generated by mixed barotropic-baroclinic instability dominates the eddy flux of zonal momentum.  The global wave converges eastward zonal momentum to the equator and deposits westward momentum at intermediate latitudes during spinup and before superrotation emerges, and the baroclinic instability ceases once superrotation is established.  A global barotropic mode of zonal wavenumber one generated by a mix of high- and low-latitude barotropic instability is responsible for maintaining superrotation in the statistically steady state.
At intermediate thermal Rossby numbers, momentum flux by the global baroclinic mode is subdominant relative to smaller baroclinic modes, and thus strong superrotation does not develop.  
\end{abstract}

\begin{article}

\section{Introduction}

Superrotation\footnote{Throughout this paper, we use the word superrotation to describe an atmosphere that has more axial angular momentum than the solid planet at the equator. This will usually occur as an angular momentum maximum at the equator, because if the atmosphere is superrotating at higher latitudes and not at the equator it will be inertially unstable. This meaning is to be contrasted with `global superrotation', which occurs when the integrated angular momentum of the atmosphere is more than it would be in solid body rotation. Global superrotation is not uncommon in planetary atmospheres, and occurs in the EarthÕs atmosphere.} is a feature of a number of planetary atmospheres, although the mechanisms that give rise to it may differ from case to case. It is a well-known result that an axisymmetric atmosphere cannot superrotate if there is a diffusive  mechanism that acts to mix angular momentum down-gradient \citep{Hide69}.
Superrotation can then only arise if there are non-axisymmetric eddy motions that act to transfer momentum upgradient and into an angular momentum maximum. Rossby waves are a potential mechanism for this: it is a property of such waves that momentum is transported in the opposite direction to their group velocity, and thus momentum converges in their source region \citep{Thompson71}. This mechanism gives rise to the eastward surface winds in midlatitudes in the Earth's atmosphere (where baroclinic instability is the source) and may lead to superrotation in two-layer models of Earth's atmosphere provided the tropical forcing is sufficiently non-axisymmetric \citep[]{Suarez_Duffy92, Saravanan93}.  Equatorial superrotation is present on Jupiter and Saturn \citep[]{Flasar_86}, requiring some form of up-gradient angular momentum transfer \citep[e.g.]{Read_86}.  Rossby waves may also be responsible for equatorial superrotation on the giant planets (where convection is likely to be the source)\citep[]{Showman07, Scott_Polvani08, Schneider_Liu09, Lian_Showman09}, although another mechanism involving equatorial Kelvin waves has been proposed \citep{Yamazaki_etal05}.  Zonal jets and equatorial superrotation on Jupiter and Saturn may alternatively result from mechanisms involving deep convection \citep[]{Heimpel_Aurnou07,Kaspi_etal09} which are unlikely to occur in shallow, terrestrial atmospheres.

Without considering the origin or nature of the eddy motion, \citet{Gierasch75} posited that large-scale eddies could be parameterized by a term in the zonal momentum equation proportional to the angular velocity gradient, and so with the same form as a molecular viscosity. The effect was further explored by \citet{Rossow_Williams79} and others in the context of superrotation on Venus. Such an eddy viscosity would tend to bring the fluid into solid body rotation, potentially leading to superrotation when coupled to a steady overturning circulation.  This effect has been invoked, at least in part, by a number of subsequent studies as being the cause of superrotation on Venus and Titan \citep{Rossow_Williams79, DelGenio_Zhou96, Hourdin_etal95, Luz_etal03, Lee_etal07}, with the proposed source of the eddies being off-equator barotropic instability.  \citet{Williams03} discussed how the eddy fluxes associated with barotropic instability can give rise to superrotation, and noted that tropical instabilities may also play a role. \citeauthor{Yamamoto_Takahashi04} \citeyearpar{Yamamoto_Takahashi04,Yamamoto_Takahashi06} found a similar effect, with the eddies in their simulations being at least partially maintained by thermal tides.  

In this study our focus is on the underlying dynamical mechanisms producing superrotation and its parametric relationship to Earth's atmospheric circulation. Thus, we use a three-dimensional model with a fairly simple treatment of the diabatic physical processes, but explore a dynamically wide parameter regime with a goal of passing from an Earth-like atmosphere to a superrotating one with a change in a single nondimensional parameter, so exposing the mechanism as clearly as possible. In section \ref{sec:moddesc} we describe the governing equations and numerical model. In section \ref{sec:nondimen} we identify the key nondimensional numbers, and in section \ref{sec:expts} we describe numerical experiments. We follow this, in section \ref{sec:dynamics} by a discussion of the dynamical processes producing superrotation and, in section \ref{sec:solarsystem}, of the relevance to planetary bodies in the Solar System. We conclude in section \ref{sec:conc}.

\begin{table}
\begin{center}
	\caption{Parameters for experiment design. A dash indicates the same value as that of the entry on its left.}
\begin{tabular}{cccc}
\Rot	 		&  		0.02   		& 		1.3			& 		10.5		  	\\ \hline
$a$	 		&  	$6.4\times10^6$m 	& 	$8\times10^5$m	& 	$2.8\times10^5$m	\\ 
$\Omega$		&  $7\times10^{-5}$s$^{-1}$& 		--			& 		--		  \\ 
$T_o$	 	&  		285 K  		& 		--			& 		--		  \\ 
$\Delta_H$	&  		0.2	   		& 		--			& 		--		  \\ 
\end{tabular}
\end{center}
\label{tab:nondimnos}
\end{table}

\pagebreak[4]
\section{Numerical Model Description} \label{sec:moddesc}

Our numerical model integrates the dry primitive equations of motion of an ideal gas on a sphere using a spectral method \citep{Gordon_Stern82} at T42 resolution (test cases were performed at higher and lower resolutions with few essential differences found).  The main differences from a comprehensive GCM are that our model has no condensate (for example water vapor for Earth, methane for Titan) and that the radiative effects are represented by a Newtonian cooling. These simplifications are severe, but allow us to focus on the dynamical aspects of the circulations of the systems.  The simplest formulation of diabatic atmospheric processes is to use linear relaxation, or Newtonian cooling, toward a specified temperature profile.  The forcing profile is fixed and a timescale, the radiative cooling time, parameterizes the time dependence.  Our approach follows  \citet{Held_Suarez94} with some differences in detail, as we now describe.

\begin{figure*}[tb]
\begin{center}
\includegraphics{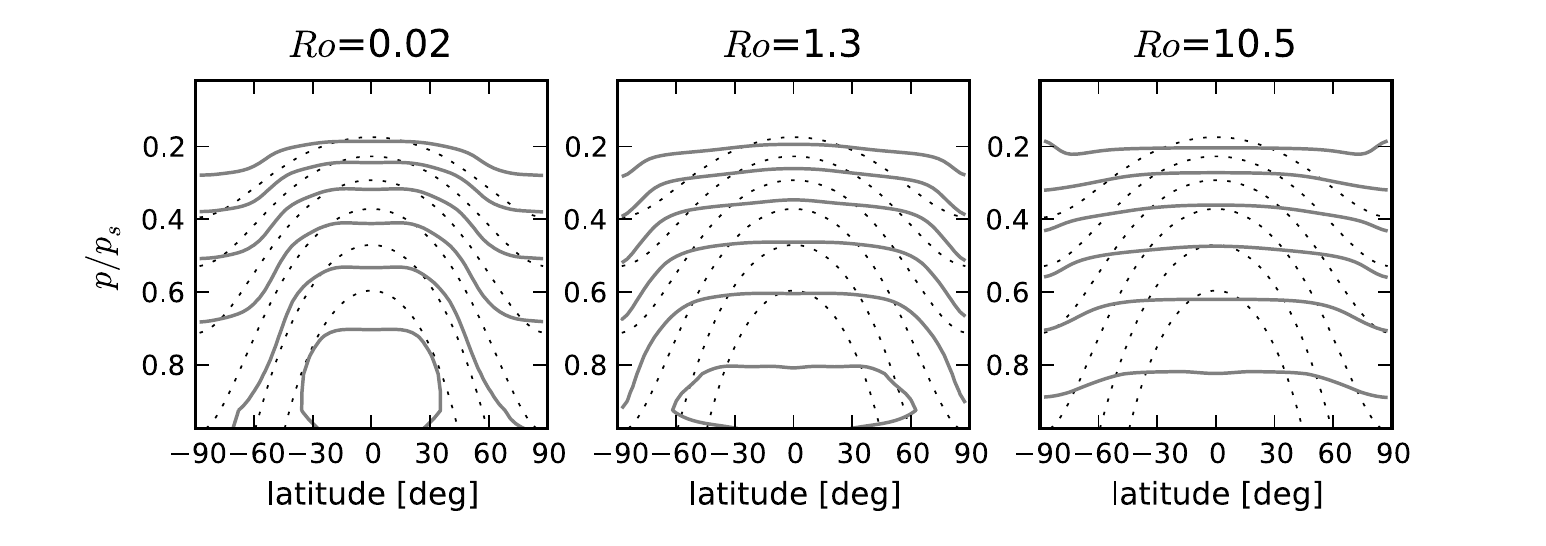}
\caption{Thermal forcing profile used for the experiments (dotted) and the zonal and time mean temperatures over the last 360 days of 1080 day integrations (gray line) for each of our three model cases.  Contours are spaced evenly at intervals of $\Delta_h/4$ from 0.7$T_o$ to $T_o$, where $T_o$ is the global mean temperature.}
\label{fig:forcing}
\end{center}
\end{figure*}

The forcing profile is assumed to be zonally symmetric, and it is assigned a fixed latitude dependence with a non-dimensional parameter, $\Delta_H$, specifying the equator-to-pole temperature gradient, $T_o = \overline{T} [1 + \Delta_H/3(1-3\sin^2\varphi)]$ with lowest-level temperature $T_o$ and global average surface temperature $\overline{T}$.  We obtain the vertical structure of the forcing by assuming a relaxation to vertical temperature profiles given by a moist adiabat with a lapse rate of 6 K/km, thus our radiative-equilibrium profile is generally stable to dry convection.  The forcing profile is capped with an isothermal stratosphere by not allowing temperatures to drop below 70\% of $T_o$.  The structure of the forcing profile is shown in \figref{fig:forcing} (dotted contours).  The specification of our radiative forcing is completed by setting the radiative relaxation time, and we use values from 40 days in the free troposphere to 4 days in the boundary layer, the top of which is fixed at $p/p_s=0.7$.

Frictional stresses exchange angular momentum between the surface and atmosphere, and we specify this mechanism as a linear Rayleigh friction in the lower atmosphere.  The Rayleigh relaxation time is taken to be 1 day at the surface and increases to the top of the boundary layer, above which there is no explicit friction except for small amount of vertical diffusion, with a diffusion coefficient of $\nu=0.01\,$m$^2$/s, that acts to smooth gridscale noise.  A fourth-order hyperviscosity is applied to dissipate energy at the grid scale.

\begin{figure*}[tb]
\begin{center}
\includegraphics{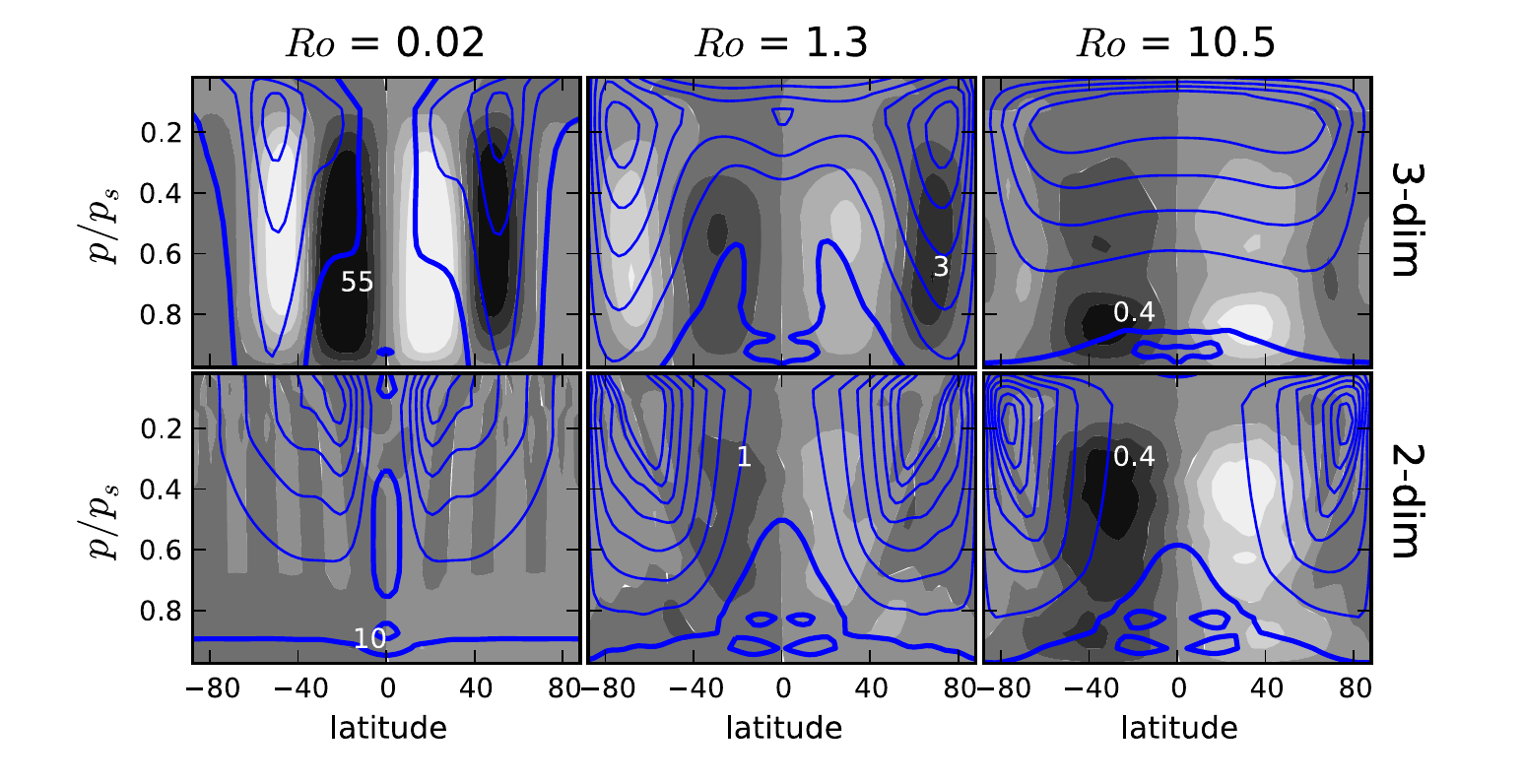}
\caption{Zonal and time mean mass streamfunction (shaded) and zonal winds (contours, zero-line bold) for \Rot= 0.02, 1.3, and 10.5.  The top row displays the full three dimensional simulations and the bottom row displays axisymmetric simulations of the same suite of parameters.  Zonal wind contours begin at 0 m/s (bold line) and are spaced by 10 m/s increments.  Mass streamfunction contour ranges are reduced by a factor of 10 for each increase in \Ro, and the maximum mass flux for each case is marked in white text [Tg/s].}
\label{fig:hadleywinds}
\end{center}
\end{figure*}

\section{Nondimensionalization and Parameter Regimes}  \label{sec:nondimen}

Given the relatively simple thermodynamic and frictional schemes described above it becomes possible to determine the nondimensional parameters that govern the behavior of the system. We begin with the primitive equations subject to Rayleigh friction and Newtonian cooling.  The horizontal momentum equations may be written in vector form \citep[Section 2.2.5]{Vallis06} as
\begin{equation}
\label{eq:dim}
	\pp \ub t + \vb \cdel \ub + \fb \cross \ub  = - \del_p \Phi - r \ub 
\end{equation}
where $\vb = u \ib+ v \jb + w \kb$, $\ub = u \ib+ v \jb$ and $\fb = f \kb = 2 \Omega \sin \varphi\, \kb$, with $(\ib,~\jb, \kb)$ being the unit vectors in the zonal, meridional and local vertical directions. The hydrostatic, thermodynamic and mass continuity equations complete the model.

Denoting non-dimensional variables with a hat, we non-dimensionalize the above equations by setting  $  
(\uhat,\vhat) = (u,v)/U; \ w = w a/(U H);
\ \Phihat = \Phi/ (2 \Omega U a); \ \that = 2 \Omega t; \ \fhat = f/(2 \Omega) = \sin \varphi$, and $(\thetahat, \That) = (\theta, T)/ (T_o \Delta_H) $; with $U$ a characteristic horizontal wind speed, 
$a$ the planetary radius, 
$H$ the scale height and $\Omega$ the rotation rate.  For the scaling of winds we use the thermal wind relation, whence $U = R T_o \Delta_H/(2\Omega a)$.  Making these substitutions into the above dimensional primitive equations we obtain 
\begin{equation}
\label{eq:nondim}
	\pp \ubhat \that + \Rot \,(\vbhat \cdel \ubhat) + \fbhat \cross \ubhat  = - \del_{\phat} \Phihat - E\ubhat \ .
\end{equation}
where the two nondimensional parameters are
\begin{subequations}
\label{eq:params}
\begin{align}
	\text{An Ekman number: } & E  =  {r \over 2 \Omega} , \\
    \text{A thermal Rossby number: }  & \Rot =  \frac U {2 \Omega a}  
           = {(R T_{0}\Delta_H)\over (2 \Omega a)^2}.
\end{align}
\end{subequations}
A third non-dimensional parameters appears in the thermodynamic equation, namely the nondimensional radiative relaxation time, $\tauhat  =  2 \Omega \tau$.  Other nondimensional parameters of the model not explicit in this derivation which we do not vary include the Prandtl number and the Rayleigh number.  The aspect ratio, $H/a$ with scale height $H$, enters implicitly through the fixed depth of the boundary layer.  Note that geostrophic balance, which is implied in the thermal wind, may not be a good approximation at high Rossby number so that although \Rot remains a valid non-dimensional parameter it may not be an accurate measure of the Rossby number itself.  The thermal Rossby number has been used as a control parameter in rotating annulus experiments \citep[e.g.,][]{Hide_58, Geisler_etal83} as well as numerical studies of Earth's atmosphere \citep[e.g.,][]{Held_Hou80}.  

In order to focus on the emergence of superrotation, we report only on the effects of the thermal Rossby number.  The Ekman number determines the strength of the drag at the bottom, and so the surface wind, and is not responsible for the qualitative difference in the regimes.  The radiative relaxation time in part determines the strength of the Hadley cell \citep[]{Held_Hou80}, but we will not vary this parameter.

\section{Numerical Experiments} \label{sec:expts}
We now describe the results of a set of numerical experiments aimed at characterizing the transition of our model atmosphere from an Earth-like case to one which superrotates.    By definition, superrotation occurs when the axial angular momentum of the atmosphere exceeds solid-body rotation, and this could in principle occur at mid-latitudes without prograde equatorial winds.   However, this would require angular momentum of the atmosphere to increase poleward, and it is likely this situation would be unstable to symmetric perturbations.  An adjustment process by symmetric eddies thus 
would eliminate the off-equatorial angular momentum maximum.  Superrotation is virtually always characterized by prograde equatorial winds and for all our experiments
prograde equatorial winds, if present, mark the atmospheric angular momentum maximum.  

It is customary to describe Titan and Venus as being in a `slowly rotating' regime compared to Earth and, for Venus, the slow rotation compared to Earth is the main factor giving it a large thermal Rossby number. Titan, however, is also substantially smaller than Earth, and this also contributes to its large thermal Rossby number.  In the present study we vary only the thermal Rossby number. In practice, because the numerical code is dimensional, we must vary a dimensional parameter and we choose to vary only the planetary radius and not the rotation rate. By making this choice the Ekman number and the nondimensional radiative relaxation time remain fixed. Both of these parameters are in fact quite different for Titan, but the simplification is useful in providing clean numerical experiments. Table 1 outlines the dimensional parameters in our experiments, focusing on three representative cases, $\Rot = 0.02,~1.3$, and 10.5. 

\subsection{The standard cases}
Global, zonal-mean diagnostics from the three standard simulations are shown in \figref{fig:forcing} and \figref{fig:hadleywinds}.  \figref{fig:forcing} displays the zonal and time mean temperature structure (gray contours) and the forcing profile (dotted contours) averaged over the final 360 days of 1080 day integrations.  Contours are spaced evenly  at intervals of $\Delta_h/4$ from $0.7 T_o$ to $T_o$ with $T_o$ the global mean surface temperature.  The most marked difference between the three cases is that as \Rot increases, the latitudinal extent of horizontally uniform temperature increases.  This feature marks the expansion of the tropical circulation pattern; in the \Rot= 10.5 case, there is only a small remnant of the original horizontal temperature gradient, now isolated to regions poleward of 60\dg N/S latitude.  Thus the state of the atmosphere becomes more barotropic as \Rot increases.  

\figref{fig:hadleywinds} shows the time and zonal mean overturning circulation (shaded) and zonal winds (contours) for our numerical experiments.  The top row displays the full three-dimensional simulations while the bottom row shows axisymmetric versions of the same three cases.  The latter were obtained by initializing the three-dimensional model without initial seed perturbations, and were verified to be very nearly axisymmetric throughout the simulations.  The color scale of the overturning circulation has been reduced by a factor of 10 between successive increases in \Rot while the scale of zonal winds is fixed (10 m/s contour spacing; zero-wind line bold).  Between $\Rot=0.02$ and $\Rot=1.3$, the Hadley cell for the three-dimensional simulations widens and substantially weakens while the indirect cell due to eddy motions in the mid latitudes is pushed poleward.  The Hadley cell in the $\Rot=10.5$ case is substantially weaker aside from a viscous cell contained in the boundary layer (pressures greater than $p/p_s=0.7$).  The mean circulation above the boundary layer in this case slants upward and poleward away from the equator.  This slant structure is characteristic of symmetric instability arising from the misalignment of contours of constant angular momentum and isentropes.

The zonal mean zonal winds in the top row of \figref{fig:hadleywinds}
transition from an Earth-like, non-superrotating regime at \Rot= 0.02 to a
fully superrotating regime at \Rot= 10.5. (Surface zonal winds in the \Rot=
10.5 case are all easterly because of the use of a Rayleigh drag in the lowest
few model layers and not just at the ground; however, the integrated torque
over the boundary layer is zero, as is required for a steady-state.) The transition between regimes occurs near
the $\Rot =1.3$ case, in which the zonal wind has Earth-like isolated
mid-latitude jets and also equatorial superrotation. This is to be expected
because, as \Ro increases past unity, the scale of baroclinic turbulence (which has a significant impact on the zonal wind structure) can no longer fit within the confines of the spherical geometry of the planet as also noted by
\citet{Williams03}. 

\begin{figure*}[htbp]
\begin{center}
\includegraphics{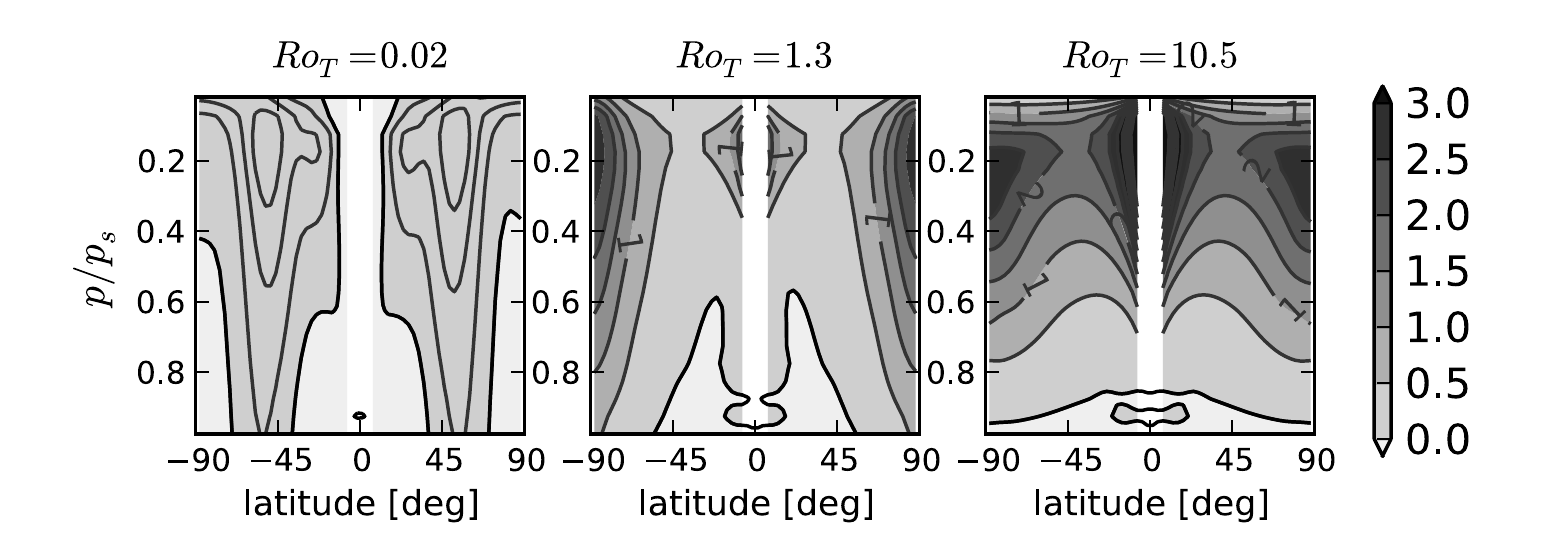}
\caption{An estimate of the local Rossby number, \Ro $\approx \overline{u}/(|f| a \cos\varphi)$ with zonal mean zonal wind $\overline{u}$, (absolute value of the) Coriolis parameter $|f|$, planetary radius $a$, and latitude $\varphi$ for our three standard cases averaged over the last 360 days of 1080 day integrations.  Shaded contours are on the same color scale in each panel, and are spaced from 0 to 3 by 0.5.  The line spacing changes for each panel:  \Rot = 0.02 lines are spaced from 0 to 0.1 by 0.02, \Rot = 1.3 are spaced from 0 to 2 by 0.5, and \Rot = 10.5 are spaced from 0 to 5 by 0.5.  The bold line marks \Ro= 0.}
\label{fig:localRo}
\end{center}
\end{figure*}

To see what causes baroclinic instabilities to become
inefficient, compare the deformation radius (which largely determines the scale
at which baroclinic cyclogenesis occurs) to the domain size. 
The deformation radius, $L_d\equiv NH/f$, does not significantly deviate from that of our thermal forcing profile in any of our simulations.  Therefore $L_d$ is essentially fixed; with typical mid-latitude values of $H\sim10$km, $f\sim\Omega$, and $N^2\sim g/T_o(g/Cp+dT_{\rm eq}/dz)\sim10^{-4}$s$^{-2}$ we find $L_d\sim1500 \km \sim a_{\rm
Earth}/4$.  Also note that, from (\ref{eq:params}b), the thermal Rossby number
\Rot is proportional to $a^{-2}$, and since in varying \Rot we are changing the
radius of our planet (while keeping the deformation radius fixed), we expect
baroclinic instability to be severely limited for cases with $a<L_d$, or in
other words for $\Ro>16\Ro_{\rm T, Earth} = 0.32$.   Figure \ref{fig:localRo} shows that in the statistically steady state in the upper troposphere, the local Rossby number, which we estimate as \Ro $\approx \overline{u}/(|f|a\cos\varphi)$ (that which is realized by the simulation), only moderately and locally exceeds unity in the \Rot = 1.3 case, indicating that baroclinic instability is still able to occur.   In the \Rot = 10.5 case the local Rossby number reaches values of \Ro = 2-5, and in this case baroclinic instabilities are inhibited because the deformation radius, and hence the scale of the instability, is becoming comparable to or larger than the planetary radius.  

The relative weakness of baroclinic
turbulence does not guarantee the development of superrotation, since there
must also be a source of momentum convergence at the equator. However, the presence of baroclinic instabilities does not of itself
preclude superrotation either, since in the absence of an equatorial critical layer,
Rossby waves generated in midlatitudes pass through the tropics without
breaking and so without decelerating the flow (see schematic in
\figref{fig:waveregimes}).  The mechanism leading to superrotation at the equator is
the topic of the Section \ref{sec:dynamics}.

\begin{figure*}[tb]
\begin{center}
\includegraphics{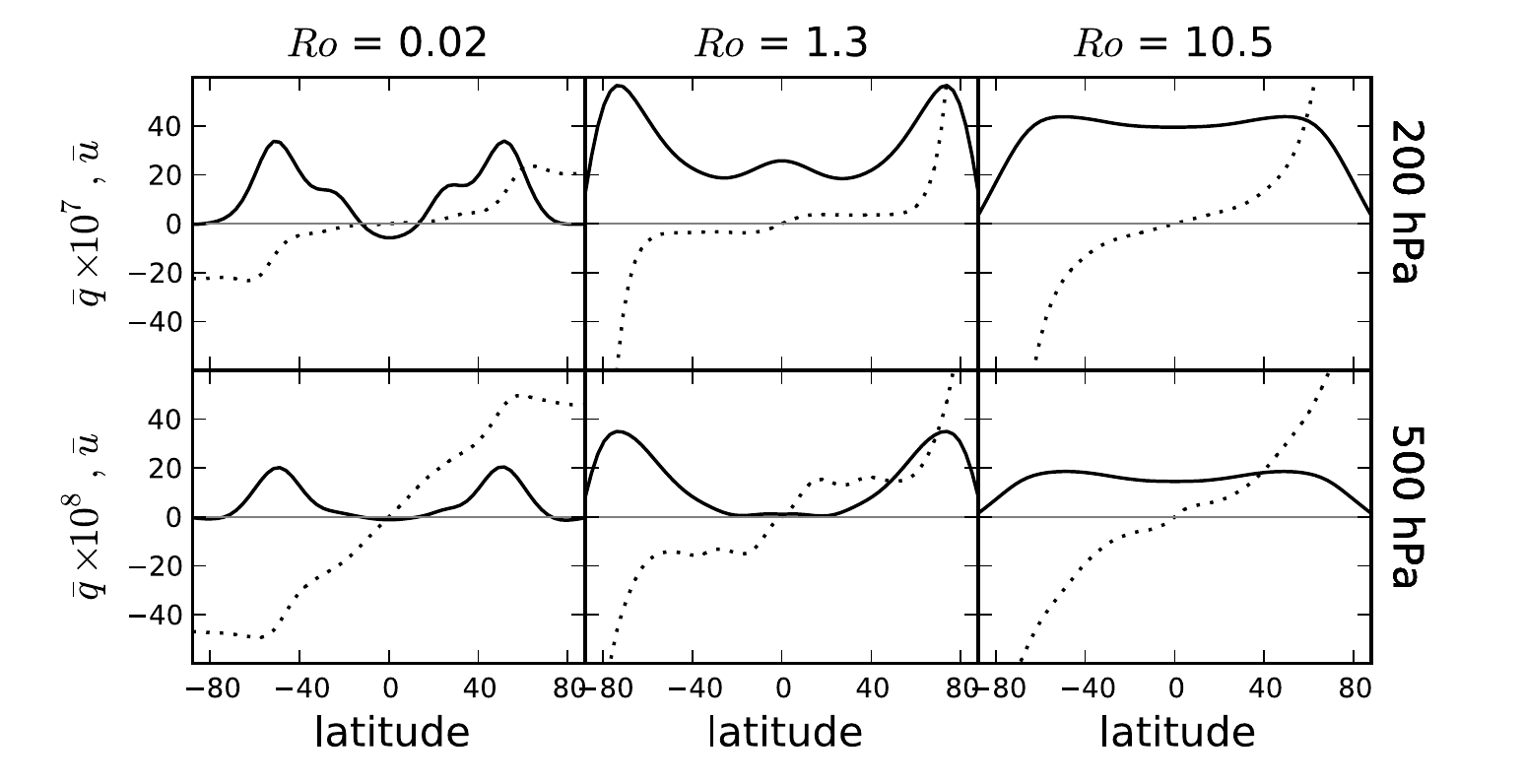}
\caption{Zonal and time mean zonal winds (solid line, m/s) and potential vorticity (dotted line) for the steady-state of \Rot= 0.02, 1.3, and 10.5 and at the 200 (top) and 500 (bottom) hPa levels.}
\label{fig:PV}
\end{center}
\end{figure*}

\figref{fig:PV} displays the zonal- and time-mean zonal winds (solid lines) and potential vorticity (PV, dotted lines).  The panels are arranged similarly to \figref{fig:hadleywinds}, with \Rot increasing from left-to-right, while the top and bottom rows display these diagnostics at 200 hPa and 500 hPa, respectively.  Especially in the middle atmosphere, the Earth-like case with \Rot= 0.02 has very non-uniform PV, while at \Rot= 1.3 the PV becomes nearly piece-wise homogeneous, forming staircase-like structures.  The \Rot= 10.5 case has very non-uniform PV at all levels. 

\begin{figure*}[tb]
\begin{center}
\includegraphics{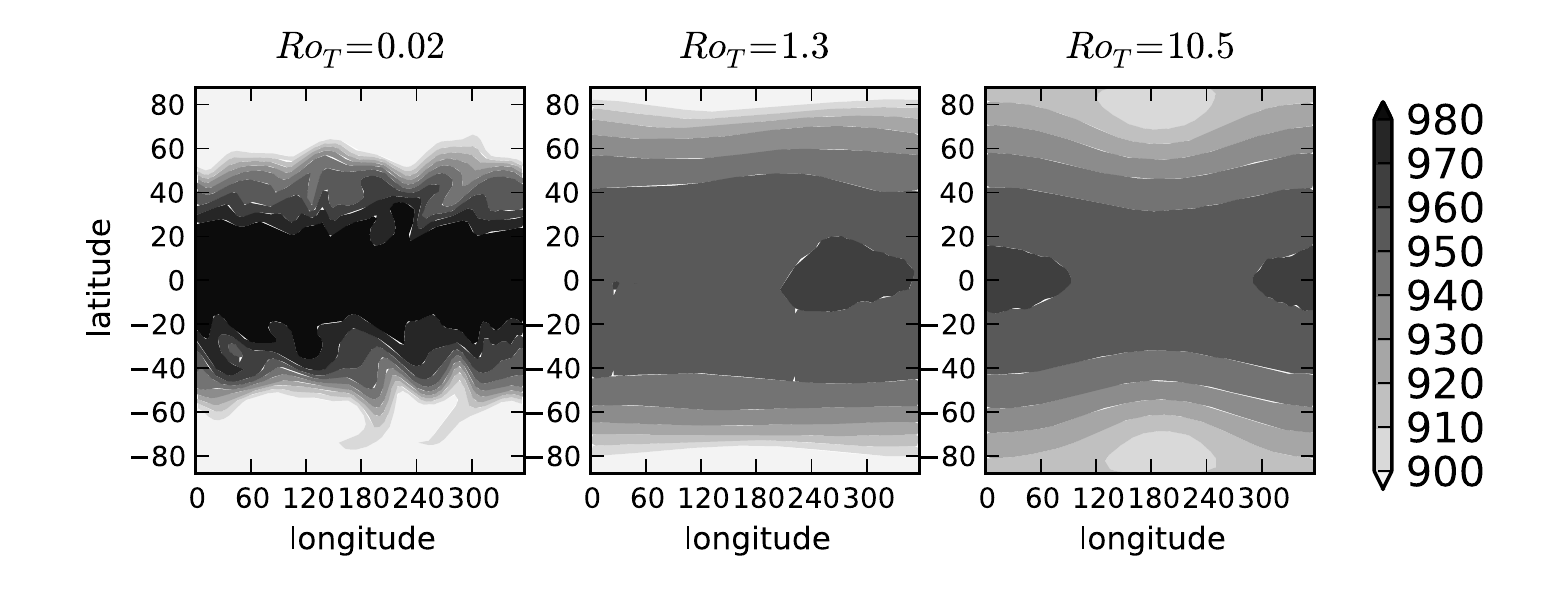}
\caption{Snapshots at day 1080 of the 400 hPa-level geopotential height in our three standard cases.  Contours are spaced evenly from 900 to 980 m in increments of 10 m.}
\label{fig:400mbGeopotential}
\end{center}
\end{figure*}

\figref{fig:400mbGeopotential} displays a snapshot on day 1080 of geopotential height on the 400 hPa level in our three standard cases.  Contours are spaced evenly from 900 to 980 m in increments of 10 m.  The \Rot=0.02 case is evidently dominated by disturbances with zonal wavenumbers $\sim$4 or larger.  These disturbances exhibit a northeast-southwest tilt in the northern hemisphere (northwest-southeast in the southern hemisphere) as is characteristic for baroclinic Rossby waves.  The \Rot=1.3 and \Rot=10.5 show clear evidence for a zonal wavenumber one disturbance with coherence at all latitudes.

\begin{figure*}[tb]
\begin{center}
\includegraphics{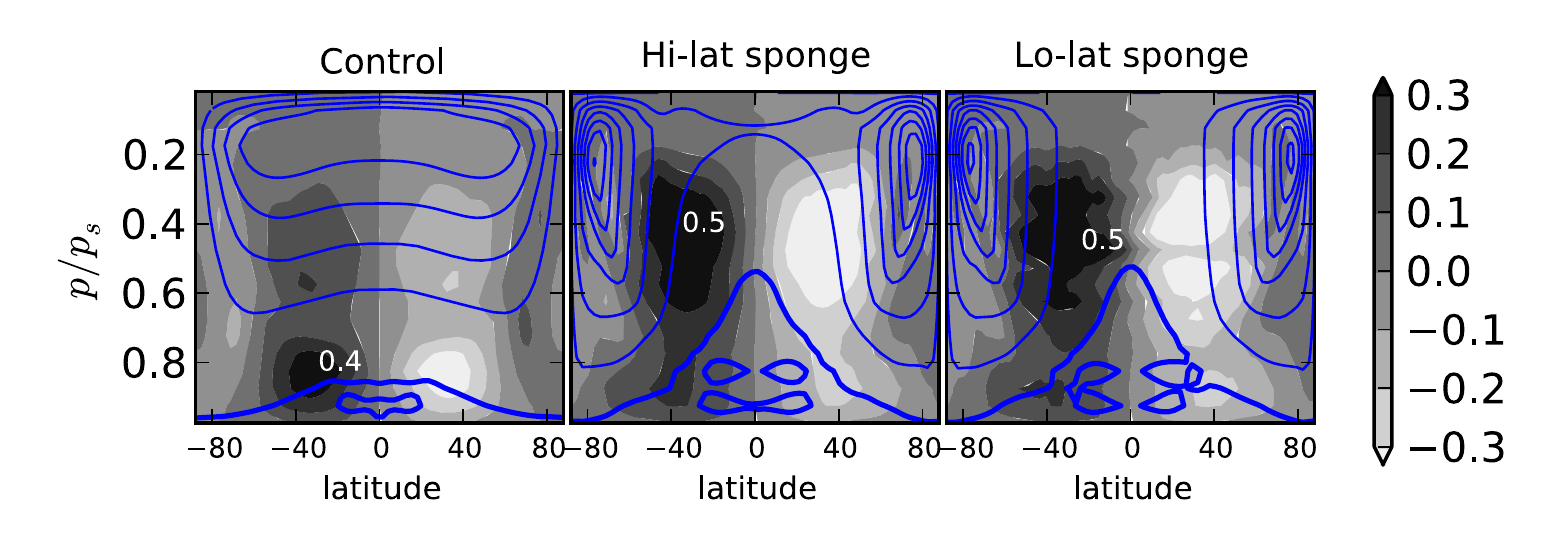}
\caption{Hadley cell and zonal mean zonal winds for test simulations of the \Rot= 10.5 case with sponge layers (contouring is as described in \figref{fig:hadleywinds} for the \Rot= 10.5 case).  The left panel displays the control case with no sponge applied.  A sponge has been applied to latitudes poleward of 45 N/S latitude in the middle panel.  The right panel has a sponge applied at the closest two grid cells to the equator, corresponding to the area between 3 N/S latitude.}
\label{fig:spongehadleywinds}
\end{center}
\end{figure*}

\subsection{The axisymmetric cases}
The axisymmetric simulations in the bottom row of \figref{fig:hadleywinds} do not develop superrotation (as they cannot in an axisymmetric flow), aside from very weak equatorial winds at high \Ro. (The very weak superrotation is due to a small amount of numerical mixing.)  It is clear from these plots that strong superrotation requires the influence of non-axisymmetric eddies.  Subtropical jets form due to the advection of high-angular-momentum air poleward from the equator, and the latitude of the jets increases with increasing \Rot as expected from axisymmetric, inviscid theory \citep[]{Held_Hou80}.

As \Rot is increased, the axisymmetric case develops a wider swath of time-dependent symmetric overturning that, in the time-mean, has the appearance of a Hadley circulation (especially in the \Rot = 10.5 case).  This axisymmetric component does not have a significant projection on the three-dimensional simulations for the \Rot = 1.3 case.  For \Rot = 10.5, the axisymmetric circulation is clearly visible in the time-mean Hadley circulation.  The similarity in the magnitude of the axisymmetric overturning in the \Rot= 10.5 case relative to the full three-dimensional simulation suggests axisymmetric theory for Hadley cells is a better approximation at higher \Rot.

\subsection{The sponge cases}
Previous studies of superrotation have often attributed equatorial eddy momentum convergence to high-latitude barotropic instabilities feeding off of horizontal shear of the jets (see e.g.~\citet{Williams03} or \citet{Hourdin_etal95}).  However, the eddies that produce the superrotation cannot be Rossby waves emanating from high-latitudes; if such waves were to radiate from a high-latitude source region and break at low latitudes, they would deposit \emph{westward} momentum thus decelerating equatorial winds.  We now demonstrate in a set of test simulations that the low-latitudes play an equally important role in genesis of the disturbances that give rise to  superrotation.  

In these test simulations, we impose a damping, or ``sponge'' on the non-axisymmetric component of all fields, but only in a select band of latitudes.  The sponge takes the form of a linear damping of the non-axisymmetric component at a timescale of 10 minutes, which results in an essentially axisymmetric state in the latitude band where the sponge is applied.  These test simulations are aimed at understanding the influence of equatorial vs. high-latitude eddies on the zonal mean zonal flow.  The first of these cases labeled ``Hi-lat sponge'' in \figref{fig:spongehadleywinds} has the sponge applied to regions poleward of 45N/S latitude.  Zonal mean zonal winds in this high-latitude sponge test case are only weakly superrotating at the equator, lending evidence of a high-latitude influence on the development of equatorial superrotation.  The second case labeled ``Lo-lat sponge'' in \figref{fig:spongehadleywinds} has the sponge applied to one grid-cell on either side of the equator, which at T42 resolution corresponds to roughly 3N/S latitude.  It is remarkable that in this low-latitude sponge case, the flow is nearly identical to the axisymmetric simulation at \textit{all} latitudes (compare with lower, right-hand panel of \figref{fig:hadleywinds}), establishing that the low-latitudes play a crucial role in the development of superrotation in the \Rot= 10.5 case.

\section{Dynamical Interpretation} \label{sec:dynamics}

We now apply several diagnostics to our simulations and give an interpretation of them.  We begin by comparing the Eliassen--Palm fluxes between the \Rot= 0.02 and \Rot= 10.5 cases.  We then use a space-time fourier analysis of the \Rot= 10.5 case to identify the disturbances that give rise to superrotation.  We end with a characterization of the disturbances contributing to the zonal mean zonal momentum and describe the mechanism by which superrotation is established and maintained.

\subsection{Eliassen-Palm fluxes}
\begin{figure*}[tb]
\begin{center}
\includegraphics{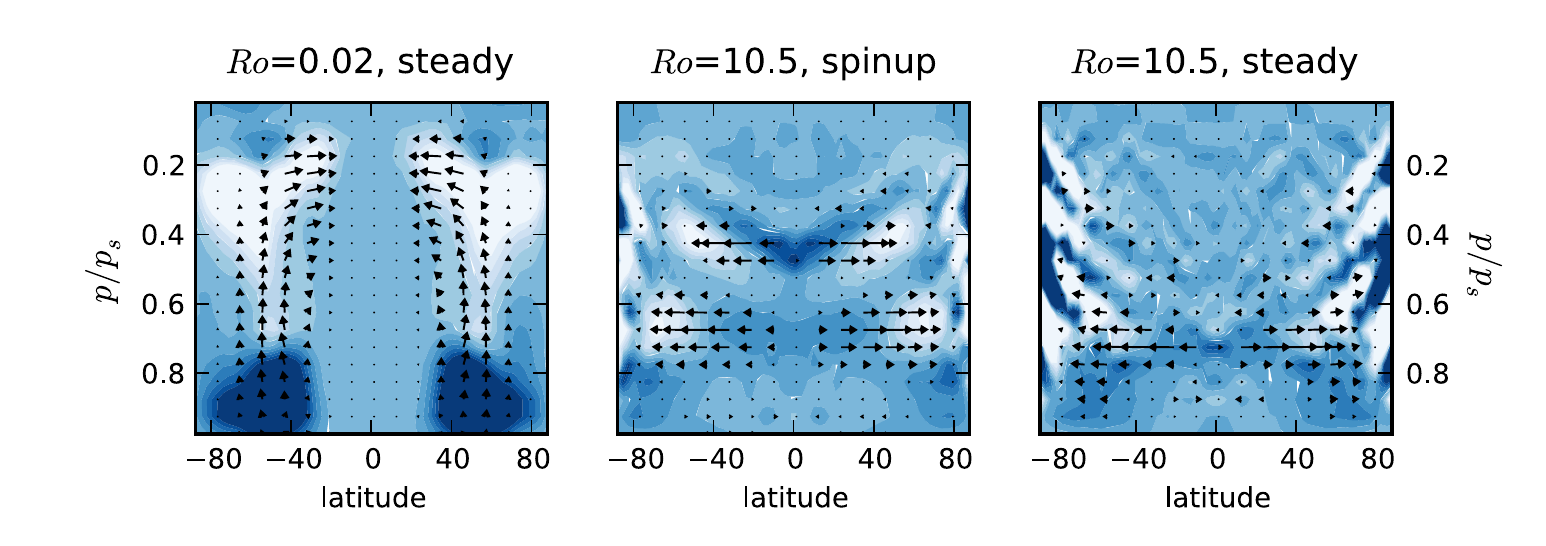}
\caption{EP fluxes and their divergences.  The left panel displays the \Rot= 0.02 (Earth-like) case after it has reached a statistically steady state (during days 1000 to 1080).  The middle panel displays the \Rot= 10.5 case (Titan-like), during active spinup, of the layer near $p/p_s=0.4$ (days 215 to 305), and the right panel displays the \Rot= 10.5 case after reaching statistically steady state (days 1000 to 1080).}
\label{fig:epflux}
\end{center}
\end{figure*}

In \figref{fig:epflux}, we show the (full ageostrophic) Eliassen-Palm (EP) flux (arrows) and its divergence (shaded) during representative times of the \Rot= 0.02 and 10.5 simulations.  The EP flux is often interpreted as a measure of the flux of wave activity, and its divergence gives the acceleration of zonal momentum due to eddy processes \citep{Edmon_etal80, Vallis06}.  In the figures, dark (light) colors indicate acceleration (deceleration) of the zonal winds.  The left panel in \figref{fig:epflux} shows the EP flux and its divergence for the Earth-like case, \Rot=  0.02, in the statistical equilibrium phase.  This has the characteristic pattern of mid-latitude baroclinic turbulence; the turbulence accelerates surface winds in mid-latitudes, decelerates the upper-level winds, and radiates Rossby waves equatorward where they are absorbed and decelerate subtropical winds. The middle panel displays the acceleration phase of the equatorial winds of the \Rot=  10.5 case, during which time there is a divergence of EP fluxes at the 400 and 700 hPa levels ($p/p_s=0.4$ and 0.7).  There is no evidence for a divergence of vertical EP fluxes.  Instead, a disturbance is contributing acceleration of equatorial zonal winds and deceleration of mid- and high-latitude zonal winds through a horizontal flux of momentum.  

Once the \Rot= 10.5 simulation is spun-up very little EP flux activity occurs in the upper, superrotating levels as seen in the right panel of \figref{fig:epflux}.  There is significant eddy activity at higher latitudes but this clearly does not directly affect the equatorial regions aloft.  The relative lack of eddy activity or mean flow (see top right panel of \figref{fig:hadleywinds}) in the upper, superrotating levels of the equator is similar to the finding of \cite{Saravanan93} that once equatorial superrotation is established the absence of equatorial torques allows the superrotation to be maintained for very long periods with little equatorial eddy activity to provide a momentum convergence.  After the spinup phase, there need be only very occasional eddy momentum fluxes to sustain superrotation, and this is the case in our \Rot = 10.5 simulation.

\begin{figure*}
\begin{center}
\includegraphics[width=5in]{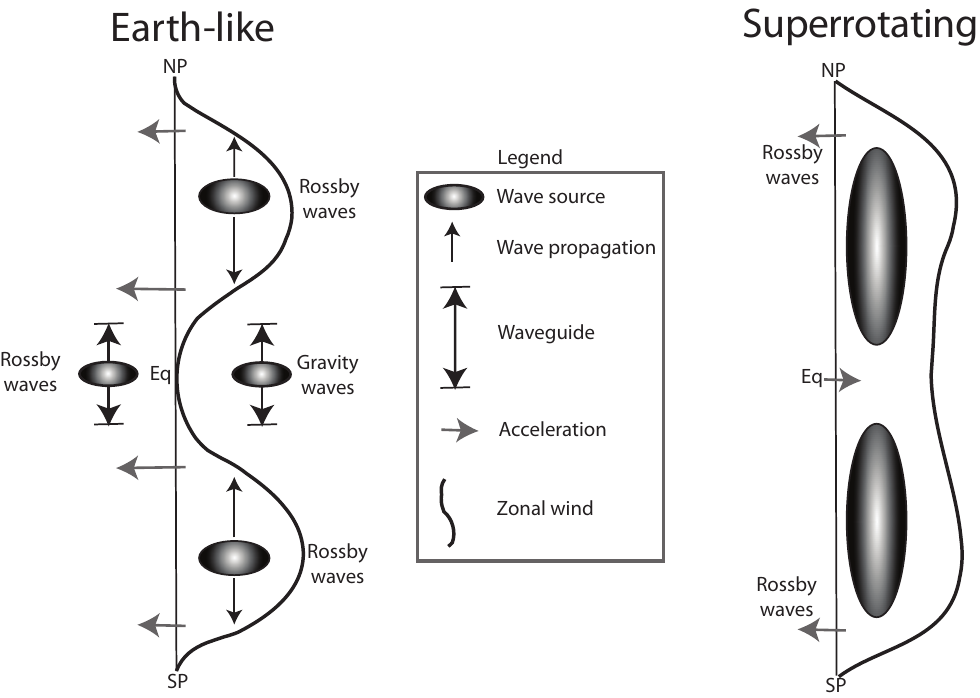}
\caption{Schematic representation of eddy-mean-flow interaction in the Earth-like (left) and Titan-like (right) regimes.  ``Wave source'' regions are intended to represent the location in latitude and characteristic phase speed of disturbances relative to the zonal wind.  In so far as dissipative effects are otherwise small, meridional propagation of these disturbances continues until either a critical layer is reached, denoted by arrows intersecting the zonal wind, or they are reflected by the vorticity gradient, denoted by arrows bounded by horizontal lines. (The generation of Rossby waves at the equator in the Earth's current atmosphere is in fact relatively weak.)}
\label{fig:waveregimes}
\end{center}
\end{figure*}

To summarize, (and see also \figref{fig:waveregimes}) in 
the Earth-like case, Rossby waves, generated in mid-latitudes by baroclinic
instability, propagate upwards and equatorwards and break, decelerating the tropical flow
and inhibiting any superrotation that might arise from other mechanisms. In the superrotating case, Rossby waves from mid- or
high-latitude regions either do not exist or do not interact with the mean flow
at low latitudes. 

\subsection{Eddy momentum flux cospectra}
We now perform spectral diagnostics of wave activity, primarily of the \Rot= 10.5 case, in order to establish those disturbances which contribute to the acceleration of the mean flow.  
Because we are interested in the influence of eddy momentum convergence on zonal mean zonal winds we compute cospectra of the horizontal flux of zonal momentum given by
\begin{equation}
\label{eq:cospec1}
	K_{n,\omega} = 2\langle {\rm Re} (U'_{n,\omega} 
	              V'^*_{n,\omega})\rangle 
\end{equation}
Here, $(U^{\prime}, V^{\prime})$ are the fourier transforms of the non-axisymmetric components of $(u,v)$, starred quantities are complex conjugates, ``Re'' refers to the real component, and angle brackets denote time averaging over a given window.  Fourier transforms are performed by direct FFT in space and time, resulting in fields gridded in zonal wavenumber, $n$, and frequency, $\omega$ \citep{Hyashi71}.  Positive (negative) frequencies represent westward- (eastward-) propagating waves.  Time averaging is accomplished in Fourier space by smoothing frequencies with a gaussian window (see \citet{Randel_Held91}).  

\subsubsection{The dominance of zonal wavenumber 1}
\begin{figure*}[tb]
\begin{center}
\includegraphics{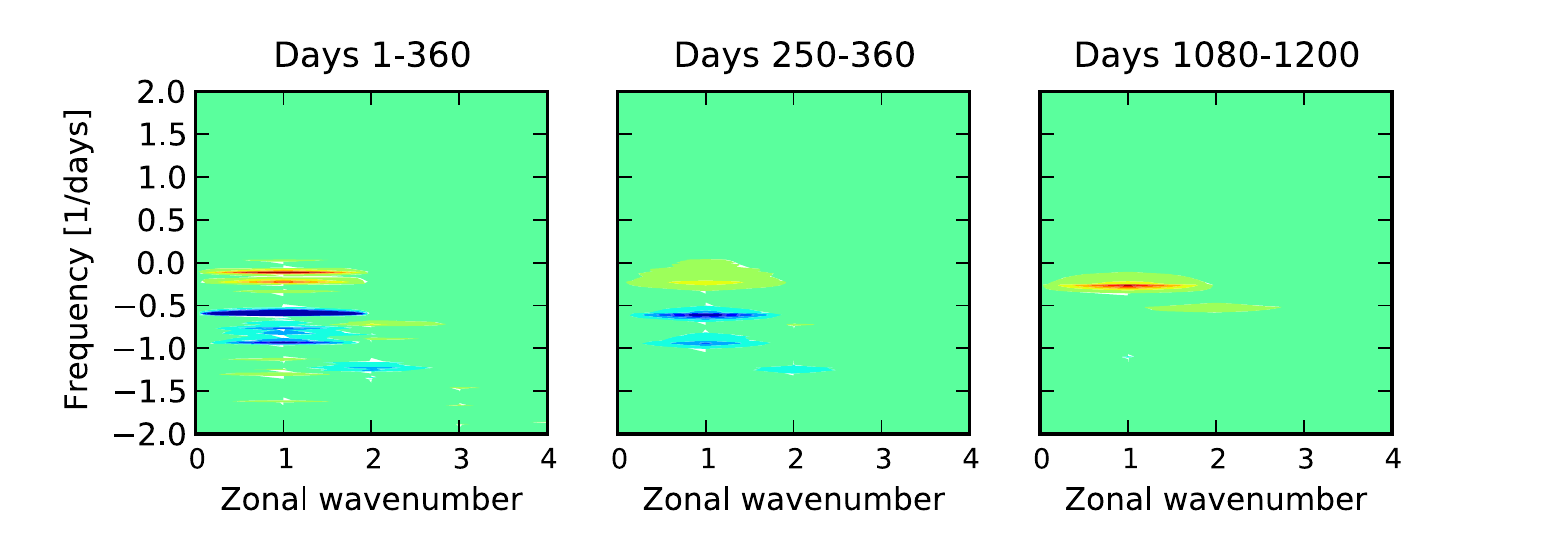}
\caption{Divergence of the eddy momentum flux cospectrum, $\partial K_{n,\omega}/\partial y$ at the 400 hPa level and at the equator for \Rot= 10.5 averaged over the days indicated.  Contour ranges are the same in each panel (their magnitudes are arbitrary).}
\label{fig:Ro10rmsUpVp}
\end{center}
\end{figure*}

In order to identify which wavenumbers interact with the zonal mean zonal wind, the left panel of \figref{fig:Ro10rmsUpVp} displays the horizontal divergence of the eddy flux cospectrum $\partial K_{n,\omega}/\partial y$ at the 400 hPa level and at the equator for the entire first model year of the \Rot= 10.5 case (negative divergence indicates eastward acceleration of the mean flow).  It is clear that zonal wavenumber one is dominating the divergence of eddy momentum fluxes, primarily at frequencies of -0.1 days$^{-1}$ and -0.6 days$^{-1}$.  

The middle panel of \figref{fig:Ro10rmsUpVp} displays the divergence of $K_{n,\omega}$ at the 400 hPa level during days 250 to 260 of the \Rot= 10.5 case, which is the most active phase of spinup.  Wavenumber one disturbances again clearly dominate the divergence of eddy momentum flux, predominantly at frequencies of $-0.25$ days$^{-1}$ and $-0.6$ days$^{-1}$.  The right panel of \figref{fig:Ro10rmsUpVp} shows the divergence of $K_{n,\omega}$ at the 400 hPa level after the model has spun up, days 1080-1200.  A wavenumber one disturbance is dominant at a frequency of -0.25 days$^{-1}$.  Eddy momentum flux is concentrated in disturbances with negative frequencies, implying eastward propagation.  However a substantial Doppler shift is present due to the strong (and latitude-dependent) zonal mean zonal wind.  We must map frequencies of the disturbances to phase speeds to understand how their phase speeds compare with the mean flow speed.  

\subsubsection{Wave-mean-flow interaction}
Linear theory predicts non-axisymmetric disturbances break and deposit their pseudomomentum into the zonal mean flow where their zonal phase speed, $c$, is roughly equal to the background zonal wind, $\overline{u}$; the breaking region is referred to as a critical layer.  For meridionally propagating disturbances, critical layers are oriented along latitude circles where the zonal mean zonal wind $\overline{u}\approx c$.  It is therefore instructive to derive the wave activity as a function of phase speed and latitude by converting the eddy momentum flux cospectra from wavenumber-frequency space to wavenumber-phase speed space and to sum the activity over wavenumber at each latitude.  The result can then be directly compared with the zonal mean zonal wind and inspected for critical layers.  

The maintenance of the mean zonal winds in the Earth-like case by eddy--mean-flow interaction is reasonably well understood and is illustrated schematically, along with a superrotating case in 
\figref{fig:waveregimes}.  In the Earth-like regime ($\Rot=0.02$), baroclinic turbulence originating in the mid-latitudes deposits westward momentum in the equatorial region, decelerating the zonal winds there.  Equatorial waves are trapped in the waveguide formed by the large-scale vorticity gradient, and so they cannot redistribute angular momentum through horizontal propagation \citep[]{Gill82}.  
If strong equatorial superrotation is present, the equatorial region becomes largely transparent to disturbances that are propagating equatorwards from higher latitudes, because the strong eastward zonal winds prevent there being a low-latitude critical layer, and so there is no eddy-damping of the eastward flow.  However, and as suggested by  \figref{fig:epflux}, to generate the superrotation in the first instance there must be a source of wave activity at or near the equator, and we now look at its spectral properties. 

\begin{figure*}[tb]
\begin{center}
\includegraphics[width=5in]{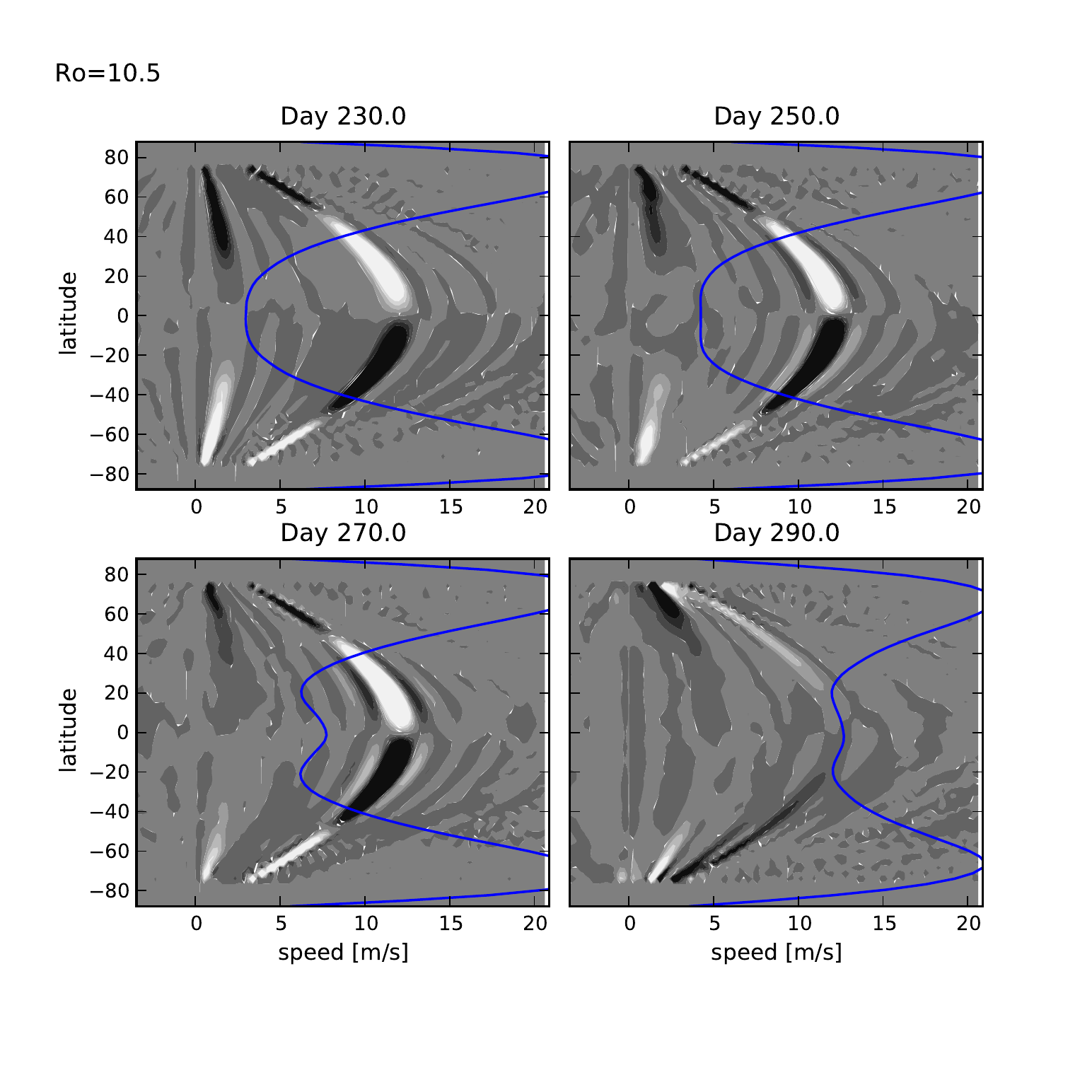}
\caption{Phase speed-latitude plot of eddy momentum flux cospectra at the 400 hPa level (shaded) and zonal mean zonal wind (line) for the $\Rot= 10.5$ case over a sequence of 20-day intervals during an active phase of spinup.  Dark (light) regions correspond to positive (negative) eddy momentum flux.}
\label{fig:Ro10phasespeedSpectra}
\end{center}
\end{figure*}

\figref{fig:Ro10phasespeedSpectra} displays the eddy momentum flux cospectra after converting frequency to phase speed, $K_{n,c_p}$ (interpolating onto a regular grid in phase speed, as in \citealp{Randel_Held91}) at the 400 hPa level and in 20-day windows centered on the day indicated (shaded, with dark or light indicating positive or negative values) with the average zonal mean zonal wind during this time overplotted (line) at a sequence of times during active spinup of the \Rot= 10.5 case.  In the first snapshot -- day 230 -- there are two global waves present with oppositely oriented momentum fluxes at low-latitudes.  The slower wave with an equatorial phase speed of a $\sim2$ m/s (corresponding to a frequency of -0.1 days$^{-1}$) is a westward-propagating Rossby wave, i.e., it is geostrophic (not shown), initiated at high-latitudes.  This disturbance decelerates equatorial winds by depositing westward (retrograde) momentum there.  The Rossby wave's influence on equatorial winds -- and in particular its deceleration -- is diminishing in this phase, since as equatorial winds accelerate the wave no longer experiences a low-latitude critical layer.  By day 250, this wave ceases to produce a momentum flux convergence at the equator.  

In each of the four snapshots of \figref{fig:Ro10phasespeedSpectra}, a faster global wave with an equatorial phase speed of 12 m/s dominates the cospectrum at low latitudes.  At the equator, this mode travels \emph{eastward} relative to the mean winds from days 230 to 270.  The momentum flux of the global wave is convergent at the equator, depositing eastward (prograde) momentum there.  The wave maintains coherence past what appears to be a critical layer at 40\dg N/S latitudes, and as a result, the global wave propagates \emph{westward} relative to the mean flow poleward of these latitudes.  The eddy momentum flux is oppositely oriented in the westward-propagating component so that the flux divergence decelerates the mean winds at 40\dg N/S latitude.  Between days 250 and 290, the zonal mean zonal wind at the equator is more than doubled by the momentum convergence of the global wave.  By day 290, mean equatorial and mid-latitude winds have accelerated beyond the phase speed of the equatorial portion of the wave and the mode responsible for equatorial acceleration no longer converges momentum to the equator.  At this point, the mean flow adjusts due to the transport of westward (retrograde) momentum to the high-latitude jets, as can be seen in the diminished winds at 70\dg N/S latitude.

\begin{figure*}[tb]
\begin{center}
\includegraphics[width=5in]{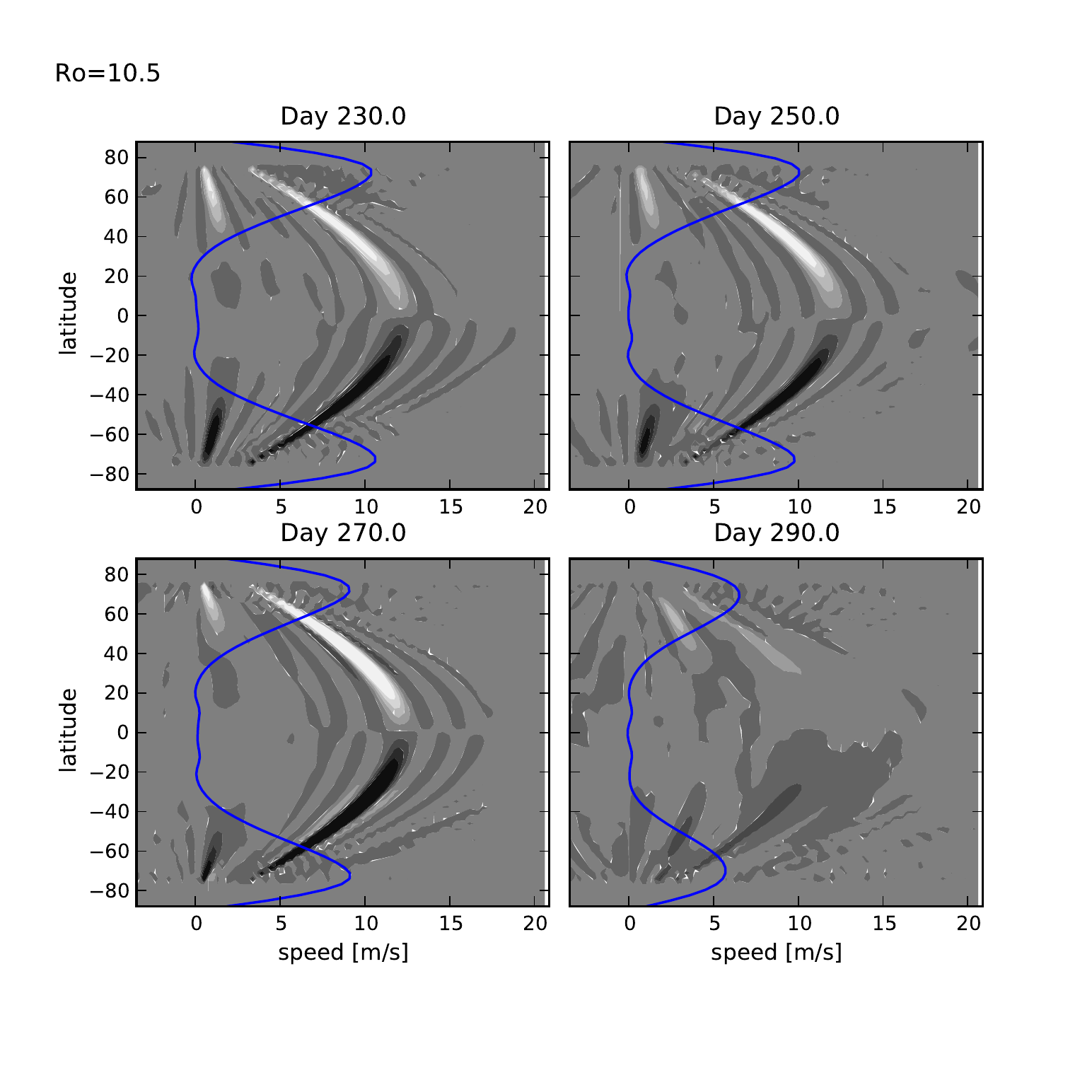}
\caption{Same as \figref{fig:Ro10phasespeedSpectra} at the 700 hPa level.}
\label{fig:Ro10phasespeedSpectra700}
\end{center}
\end{figure*}

\figref{fig:Ro10phasespeedSpectra700} displays the eddy momentum flux cospectra at the 700 hPa level for the same sequence of times during spinup shown in \figref{fig:Ro10phasespeedSpectra}.  The same global disturbances dominate the eddy momentum fluxes at this level, with some important differences.  The faster wave converges momentum to the equator as it does at the 400 hPa level, but it does so globally without a reversal in the direction of momentum flux (as occurs at the 400 hPa level poleward of 40 N/S latitudes).  Momentum fluxes at the 700 hPa level due to the slower wave are also oppositely oriented relative to those at the 400 hPa level.  The 700 hPa level resides at the top of our boundary layer, and a frictional torque prevents equatorial superrotation from developing despite strong eddy momentum flux convergence there.  

\begin{figure*}[tb]
\begin{center}
\includegraphics[width=5in]{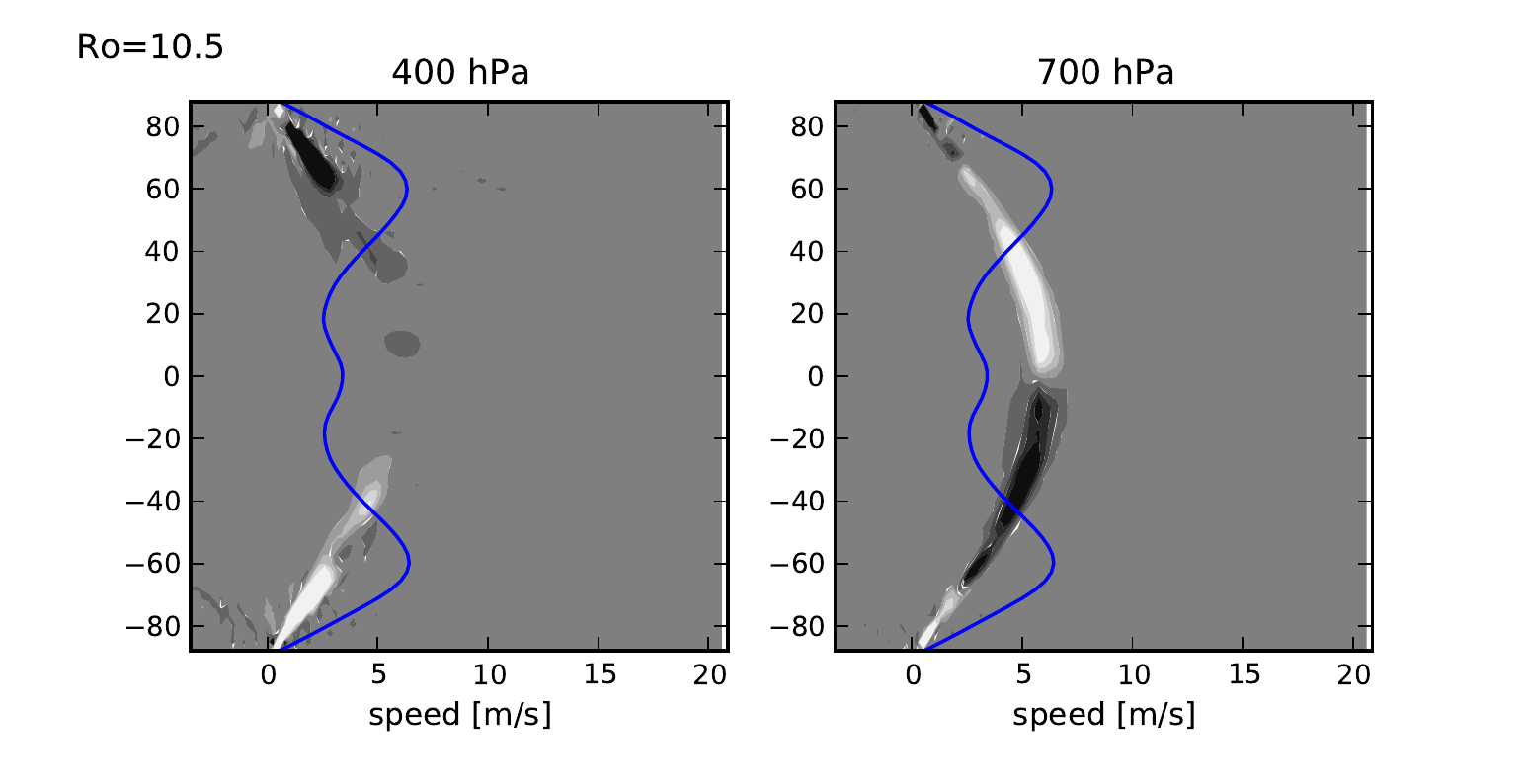}
\caption{Eddy momentum flux cospectra at 400 (left) and 700 (right) hPa in the statistically steady state.  Note that the global wave now has an equatorial phase speed of $\sim$5 m/s.}
\label{fig:Ro10PhaseSpeedCompare}
\end{center}
\end{figure*}

\figref{fig:Ro10PhaseSpeedCompare} displays the 400 and 700 hPa (left and right, respectively) eddy momentum flux cospectra in the statistically steady state, i.e., after superrotation is established.  A single, global wave dominates the cospectrum at a slower phase speed, $\sim$5 m/s at the equator, than occurs during active spinup.  The global wave at the 700 hPa level deposits prograde momentum at the equator.  The equatorial eddy momentum flux is not well-organized at the 400 hPa level during this phase.  However, the free atmosphere at this level does not directly experience frictional torques from boundary layer processes (which are confined to $p>700$ hPa), has no significant zonal mean overturning circulation (as can be seen in the upper left panel of \figref{fig:hadleywinds}), and does not experience a vertical divergence of eddy activity (as can be seen in the left panel of \figref{fig:epflux}).  Consequently, the relatively weak and disorganized horizontal eddy momentum flux convergence is able to maintain the superrotating state.  Section \ref{sec:barotropic} contains further discussion of the mechanism maintaining superrotation.

\subsection{Structure of the global wave producing \newline superrotation}

\begin{figure*}[tb]
\begin{center}
\includegraphics{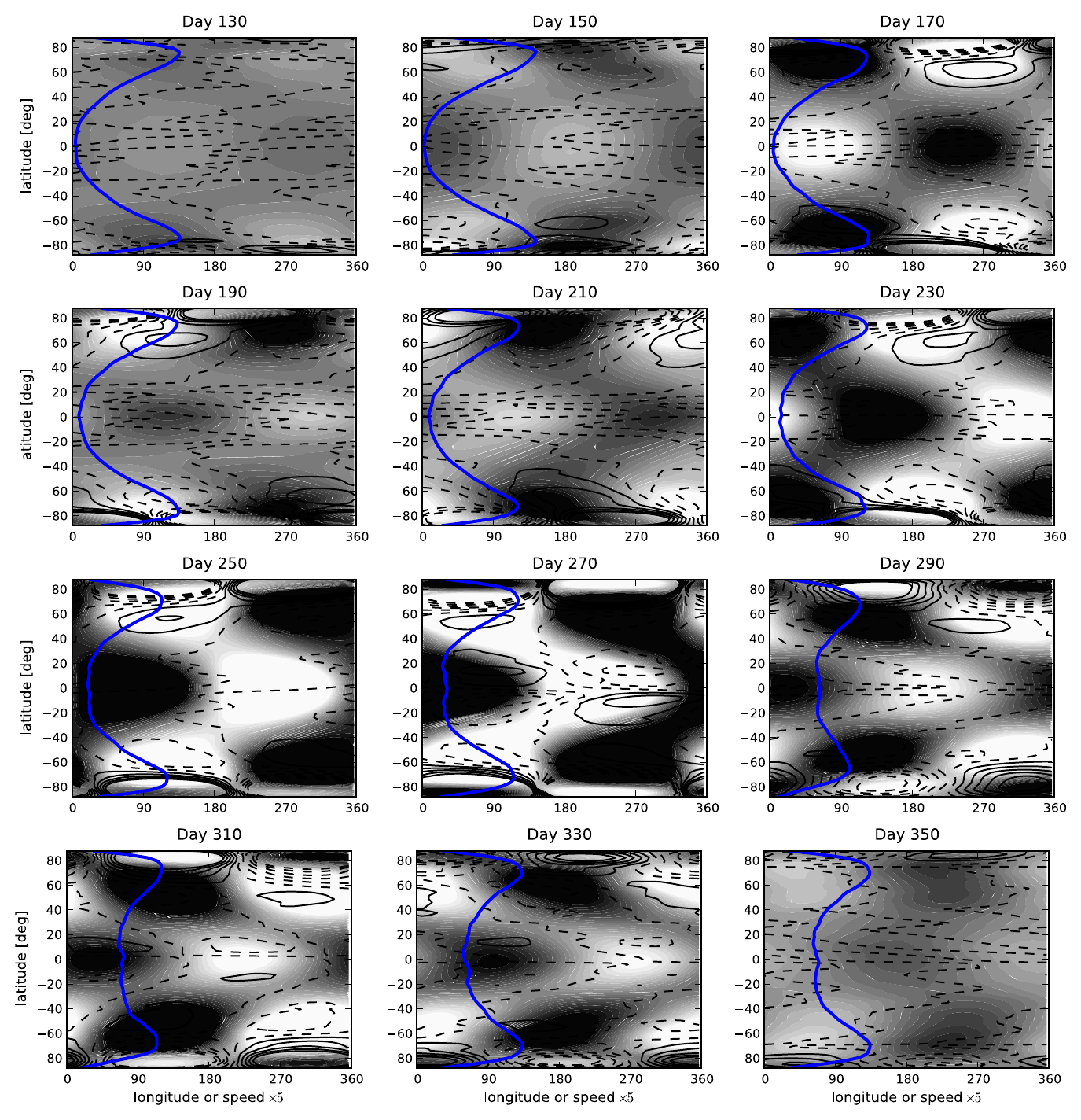}
\caption{Sequence of snapshots of zonal wavenumber one, $-0.6$ days$^{-1}$ frequency geopotential (shaded) and potential vorticity (contours) with zonal mean zonal wind multiplied by a factor of 5 overplotted (line) at the 400 hPa level for the \Rot= 10.5 case. During the course of the transition to superrotation, particularly days 210--310, note how the disturbance remains coherent in latitude even as it propagates eastward (as is evident in \figref{fig:Ro10phasespeedSpectra}) and that it leans into the horizontal shear.  Note that the wave travels once around the globe in $1/0.6\sim$1.7 days, and therefore eastward propagation of the pattern is not evident due to aliasing.}
\end{center}
\label{fig:globalwave2}
\end{figure*}

The spectral diagnostics in Figures \ref{fig:Ro10rmsUpVp}--\ref{fig:Ro10PhaseSpeedCompare} together reveal the disturbance responsible for equatorial momentum convergence during spinup has zonal wavenumber one and frequency -0.6 days$^{-1}$.  
\figref{fig:Ro10phasespeedSpectra} also shows the disturbance propagates both eastward and
westward relative to the zonal mean zonal wind, depending on the latitude, but,
importantly, the disturbance retains a coherent latitudinal structure. Figure
\ref{fig:globalwave2} displays a sequence of snapshots of the 400 hPa geopotential (shaded)
and potential vorticity (PV, contours) during spinup filtered for zonal wavenumber one and
frequency $-0.6$ days$^{-1}$ (the zonal mean zonal wind multiplied by a factor
of 5 is also displayed).\footnote{Filtering is performed by direct FFT in longitude and time, selecting the desired wavenumber and frequency, and performing a reverse FFT on the filtered spectrum.}  During days 130 to 210, the equatorial (eastward-propagating) component is beginning to phase with the high-latitude (westward-propagating) component.  Both components strengthen during days 230 to 290, during which time the equatorial wind speeds double in magnitude due to horizontal eddy momentum convergence there.  The components then weaken to roughly their strength at day 130 and de-phase, thus ending a period of rapid  acceleration of equatorial zonal winds.

Near the equator, the global wave at times seems to resemble an equatorial Kelvin (gravity) wave with a
characteristic chevron shape. However, the phase speed of the equatorial
portion (12 m/s) does not correspond to typical gravity wave speeds, the
gravest of which would have $c>$100 m/s.
Poleward of 40 N/S latitudes (the location of the critical layer in Figure
\ref{fig:Ro10phasespeedSpectra}) the wave propagates westward relative
to the mean flow but it remains anti-phased with the equatorial portion as it evolves.

\begin{figure*}[tb]
\begin{center}
\includegraphics[scale=0.9]{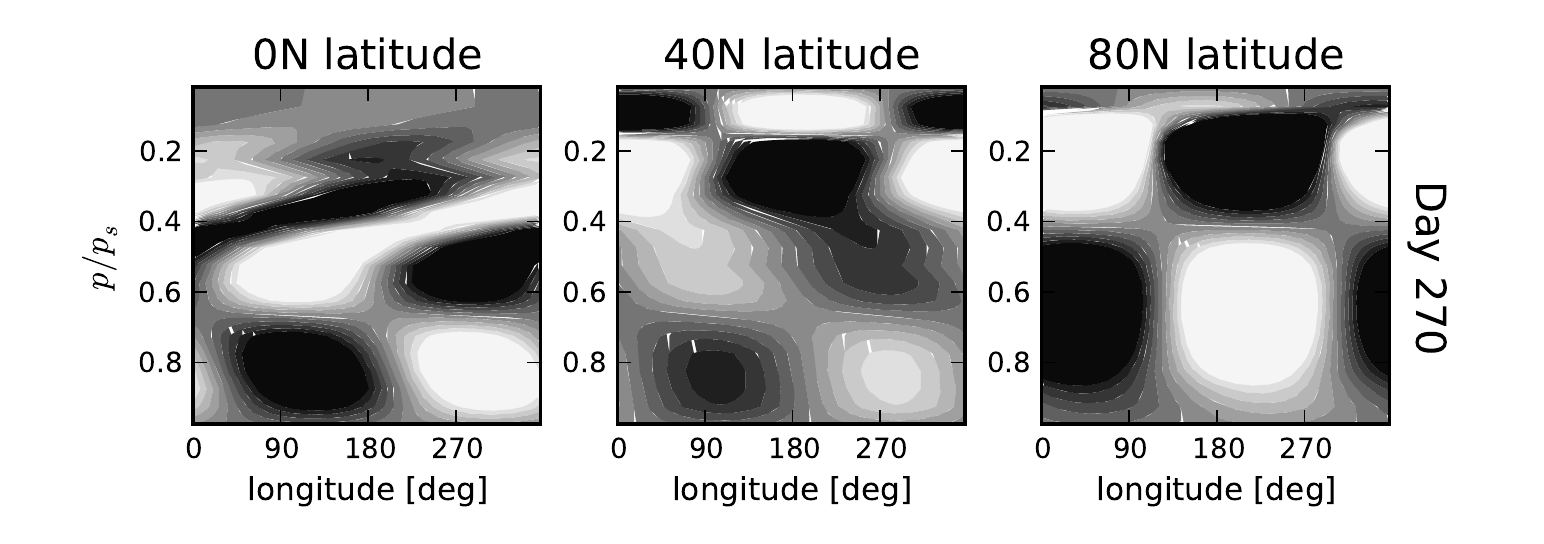}
\includegraphics[scale=0.9]{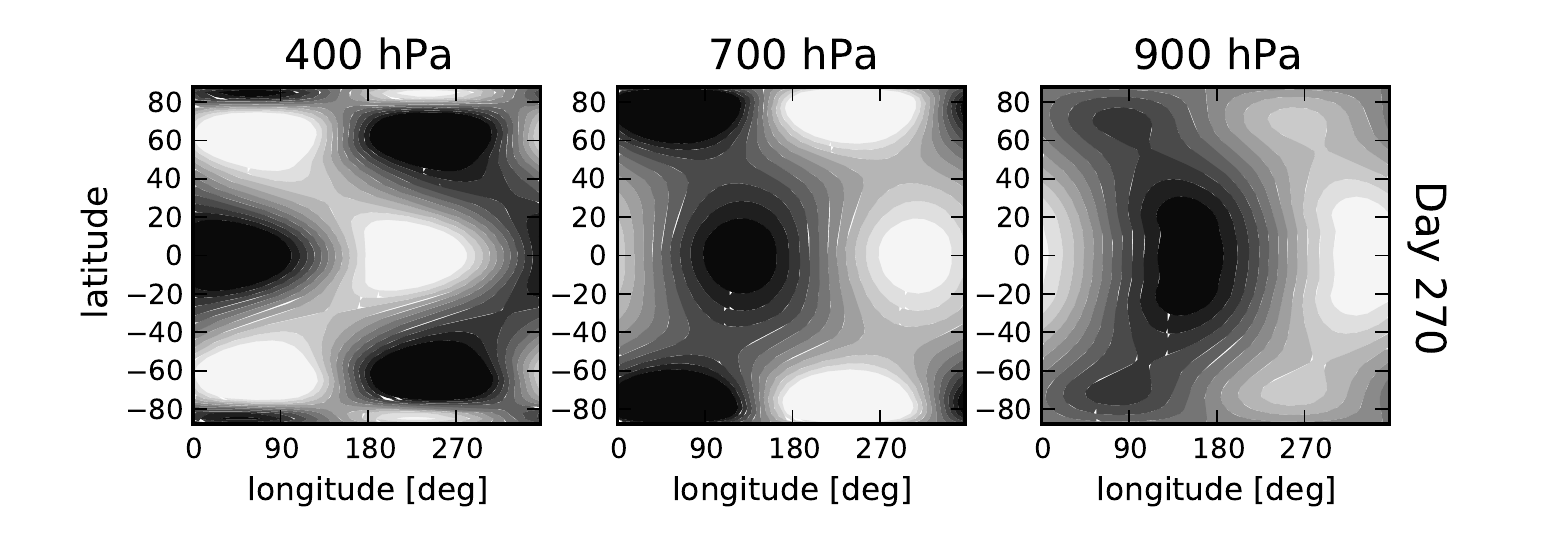}
\caption{Vertical and horizontal slices of zonally asymmetric geopotential filtered for zonal wavenumber 1 and frequency $-0.6$ days$^{-1}$ (shaded) during spinup (day 270) at the latitudes or levels indicated.  Contours are spaced evenly from $-0.5$ to 0.5 m in increments of 0.1 m.}
\label{fig:verticalstructure}
\end{center}
\end{figure*}

\figref{fig:verticalstructure} displays the geopotential height anomaly associated with wavenumber 1 disturbances contributing to eddy momentum flux on day 270 (during spinup) with a frequency of -0.6 days$^{-1}$.  Contours are spaced evenly from -0.5 to 0.5 m in increments of 0.1 m.  The top row of panels show the pressure-longitude cross-section at three latitudes (the pattern moves toward the right in these figures).  The disturbance is evidently the first baroclinic mode, as can be most clearly seen in the vertical slice at 80N latitude (top-right panel).  At lower latitudes, the vertical structure is more complicated which may be the result of shearing by the zonal mean zonal wind.  The bottom row displays latitude-longitude cross-sections of the wave at the pressure levels indicated (for reference, the lower left panel corresponds to the level displayed in \figref{fig:globalwave2}).  At 900 hPa, the wave appears to be coherent across all latitudes with a slight northwest-southeast (northeast-southwest) tilt in the northern (southern) hemisphere.  This tilting is more pronounced at 700 hPa, and still more so at 400 hPa.

\begin{figure*}[tb]
\begin{center}
\includegraphics[scale=0.9]{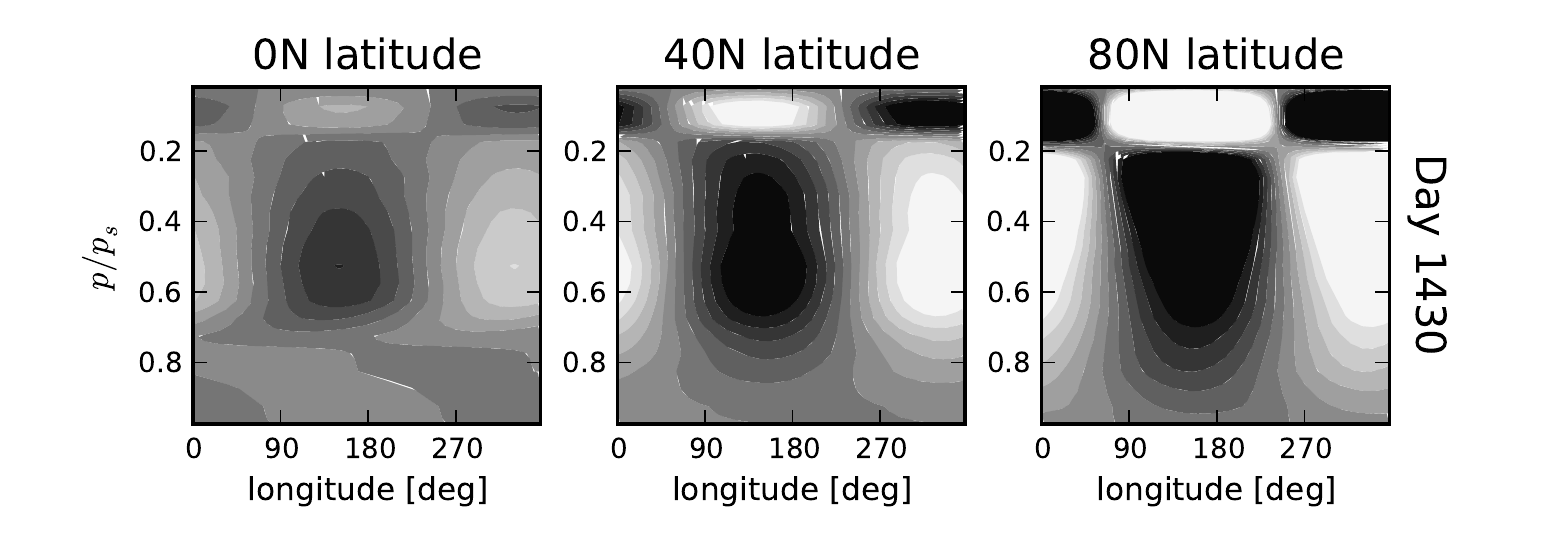}
\includegraphics[scale=0.9]{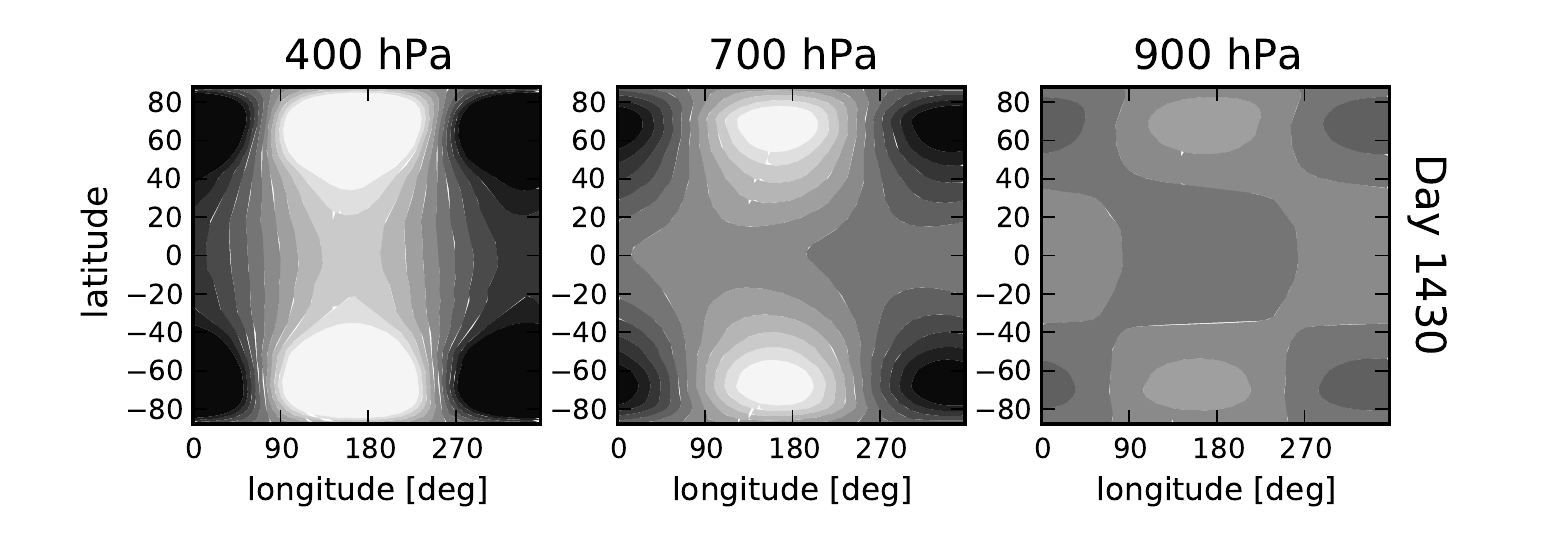}
\caption{Same as in \figref{fig:verticalstructure} for the global wave, now with frequency $-0.25$ days$^{-1}$, in the statistically steady-state (day 1430).}
\label{fig:steadyverticalstructure}
\end{center}
\end{figure*}

\figref{fig:steadyverticalstructure} displays the geopotential height anomaly associated with wavenumber 1 disturbances contributing to eddy momentum flux in the statistically steady-state (day 1430 is shown).  This disturbance, with a frequency of -0.25 days$^{-1}$, is clearly barotropic aside from moderate tilting that is evident in the frictional boundary layer.  Its horizontal structure is very uniform across all latitudes except for in the frictional boundary layer where the contours tilt northwest-southeast (northeast-southwest), in the same manner they did for the wave contributing to spinup in \figref{fig:verticalstructure}.  

The cause of the northwest-southeast (northeast-southwest) tilting and its influence on the zonal mean zonal wind is now discussed.

\subsection{Structure of the zonal mean potential vorticity}
\begin{figure*}[tb]
\begin{center}
\includegraphics{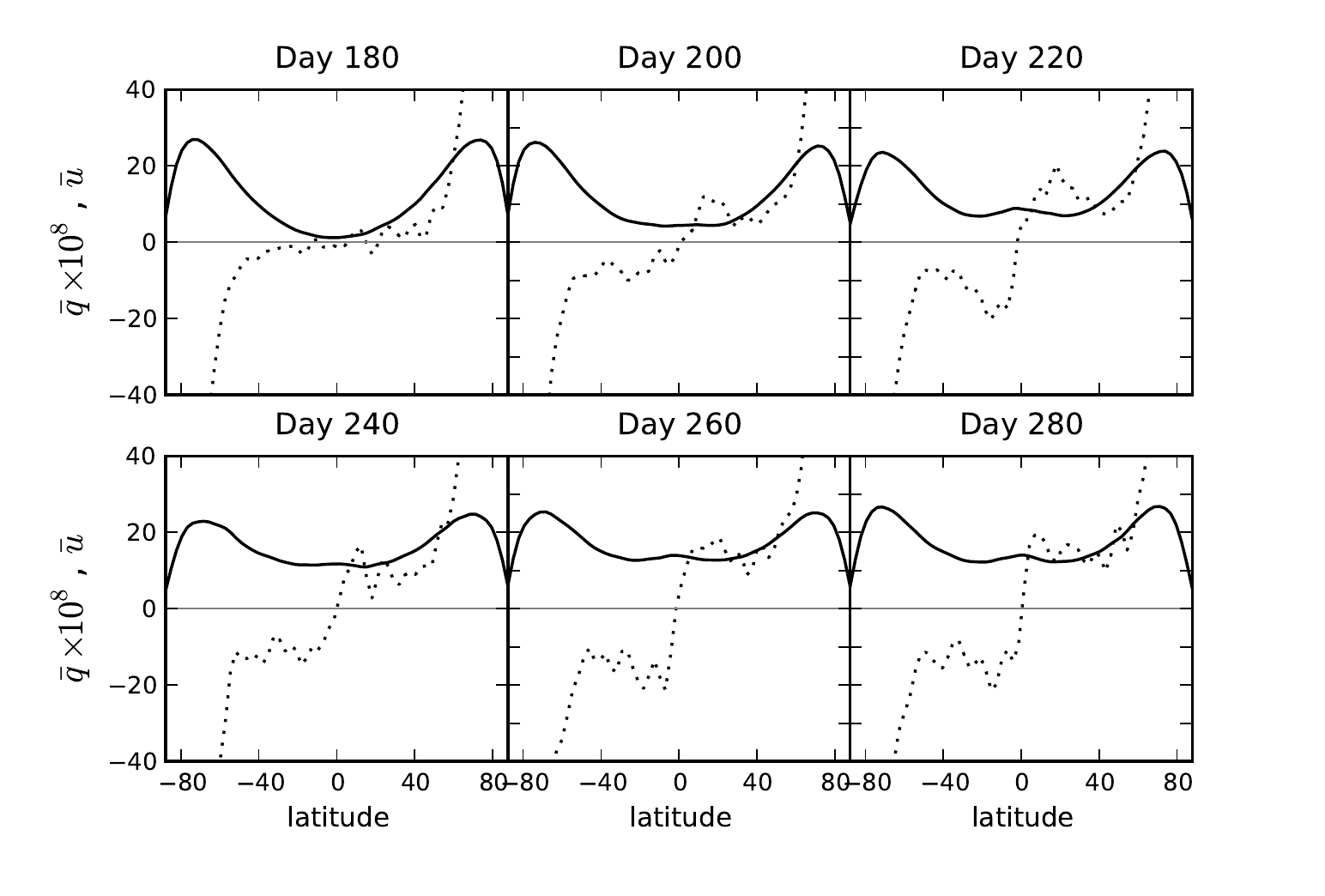}
\caption{Same as in \figref{fig:PV} for snapshots during spinup (on days indicated) of the \Rot= 10.5 case at 400 hPa.}
\label{fig:PVsnapshots}
\end{center}
\end{figure*}

The development of the instability can be seen in the time-evolution of zonal mean PV and zonal winds in \figref{fig:PVsnapshots}, shown at the 400 hPa level of the \Rot=10.5 case on the days indicated.  A necessary condition for bartotropic instability is that the meridional gradient of zonal mean PV, $\partial \overline{q}/\partial y$, change sign in the horizontal.  At day 180 of \figref{fig:PVsnapshots}, mean equatorial winds are very calm and the equatorial PV is nearly zero aside from small perturbations.  Zero PV is an indication of an angular-momentum-conserving overturning circulation during the initial, nearly axisymmetric phase of the simulation.  The PV gradient at the equator has steepened by day 200, which introduces areas of negative PV gradients near 20\dg N/S latitude.  Equatorial winds accelerate during this time, which is suggestive of a link between a large-scale barotropic instability and the emergence of superrotation.  

\begin{figure*}[tb]
\begin{center}
\includegraphics[scale=0.95]{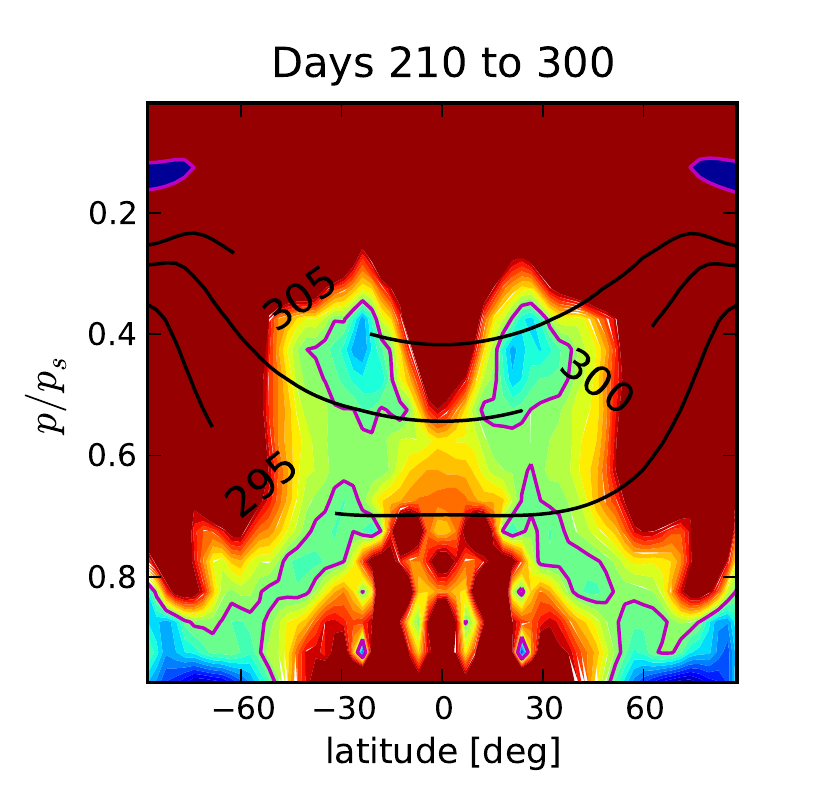}
\includegraphics[scale=0.95]{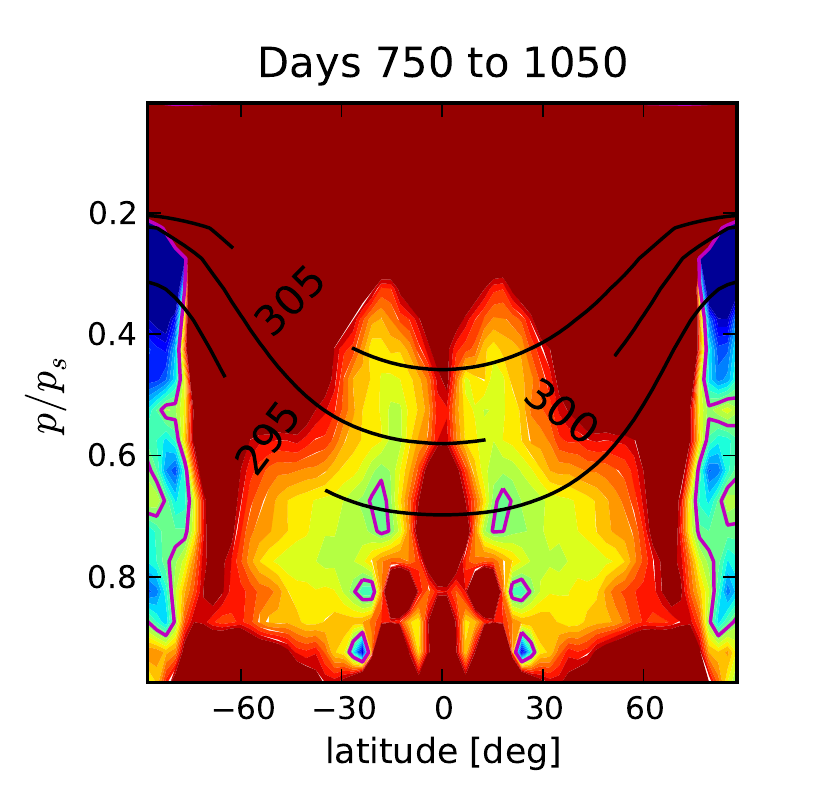}
\caption{\emph{(Top row)} Zonal- and time-mean PV gradient, $\partial \overline{q}/\partial y$, during an active phase of spinup (left panel) and in the steady-state (right panel) for the \Rot= 10.5 case.  The zero PV gradient line is marked with a magenta line, with warm (cool) colors indicating positive (negative) PV gradient.  The 295K, 300K, and 305K isentropes are overplotted in black.  \emph{(Bottom Row)} Same on isentropes.  White lines indicate areas of large, positive PV gradient.}
\label{fig:PVgradients}
\end{center}
\end{figure*}

The zonal mean PV gradient in the \Rot= 10.5 case during spinup is displayed in left panel of \figref{fig:PVgradients}.   The zero PV gradient line is marked with a magenta contour, and three isotherms are over-plotted in black.  There is clear evidence for a PV gradient reversal near the surface and at high latitudes, indicating a role for baroclinic instability during spinup.  The global (first) baroclinic mode responsible for spinup (\figref{fig:verticalstructure}) is associated with this high-latitude baroclinic instability.  Being global in scale, the disturbance is unable to propagate either vertically or horizontally and thus our notions based on the propagation and breaking of linear waves is insufficient to explain its interaction with the mean flow.  

The PV gradient also reverses sign on either side of the equator at the 400 hPa level and at 30 N/S latitudes.  \emph{Eastward}-propagating Rossby waves initiated in these regions of PV reversal span the equator and interact with one another.  This accounts for the slow, eastward propagation (relative to the gravity wave phase speed) of the equatorial portion of the global wave.  The eastward propagation of the equatorial portion together with an eastward, horizontal zonal wind shear in zonal winds allow phase-locking between high-latitude (westward) and low-latitude (eastward) portions of the global wave (as is a characteristic of barotropic instability -- see Section \ref{sec:barotropic}).

The zonal mean PV gradient in the \Rot= 10.5 case in the statistically steady state is displayed in right panel of \figref{fig:PVgradients}.  The regions of negative PV gradients at the high-latitude surface are no longer present, indicating an absence of baroclinic instability.  The high-latitude PV gradient reverses sign in the horizontal, indicating barotropic instability is associated with the barotropic mode seen during the steady-state (\figref{fig:steadyverticalstructure}).  The middle troposphere at low latitudes no longer has PV gradient reversals that were present during the spinup phase, although there are small regions of negative PV gradient confined to the boundary layer (pressures greater than 700 hPa).  

The PV diagnostics in Figs. \ref{fig:PVsnapshots} \& \ref{fig:PVgradients} together suggest a role for baroclinc \emph{and} barotropic instabilities in the emergence and maintenance of superrotation in our simulations.  It is unclear from our experiments whether both instabilities are necessary for the transition to occur.  The mechanism giving rise to a convergence of eddy momentum flux at the equator is very similar in character to barotropic instability, which we now discuss in more detail.

\subsection{Barotropic instability}\label{sec:barotropic}
\begin{figure}[tb]
\begin{center}
\includegraphics[scale=0.4]{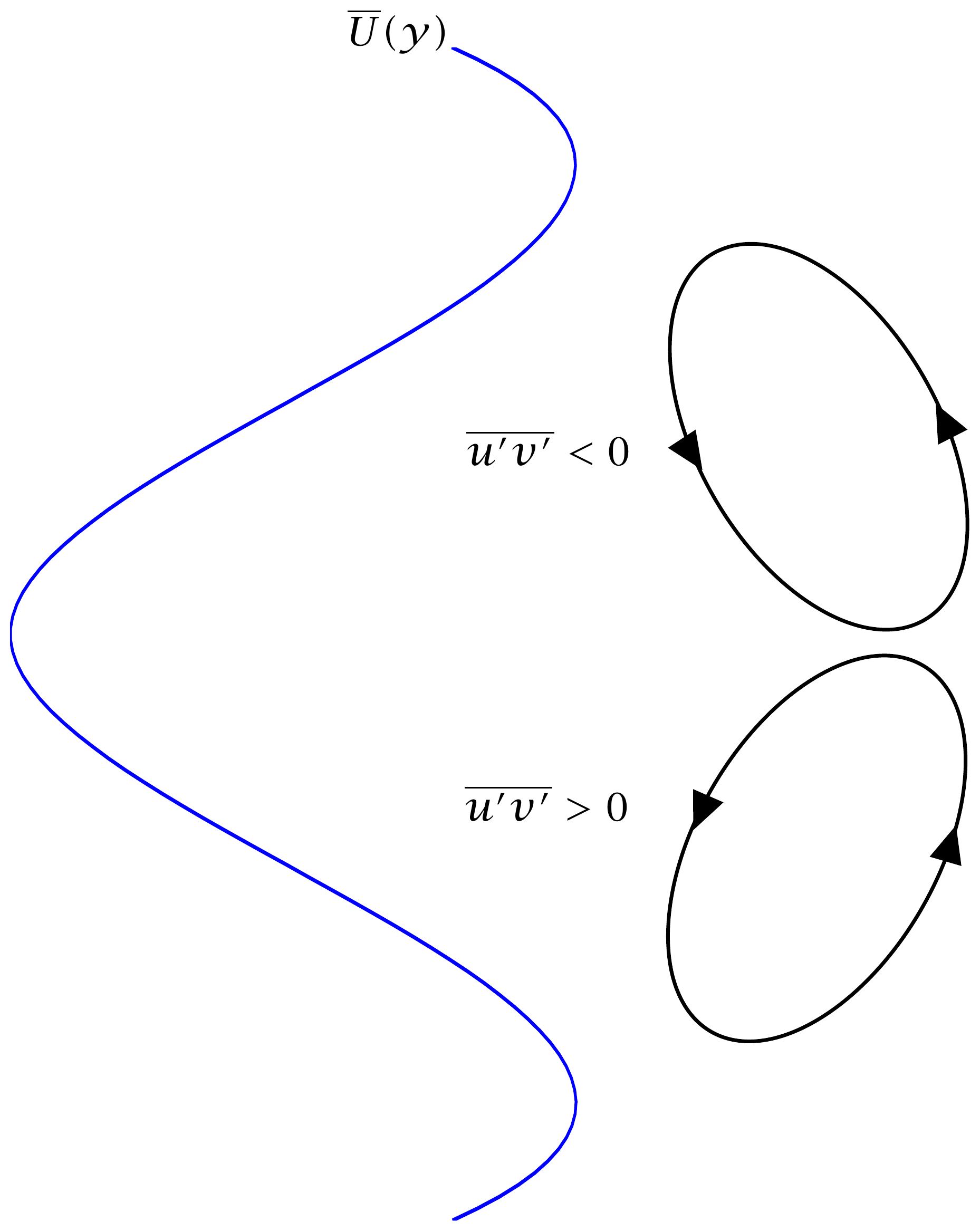}
\caption{Schematic of barotropic instability during spinup of the \Rot= 10.5 case.  The meridional structure of the zonal mean zonal wind is shown as the line.  Interacting edge waves form circulation patterns that slant into the shear of the zonal winds.  The resulting Reynolds stresses converge momentum to the equator.}
\label{fig:barotropic}
\end{center}
\end{figure}

The time-evolution of geopotential and potential vorticity shown in
\figref{fig:globalwave2} and \figref{fig:PVsnapshots} and the reversals of PV gradients present in \figref{fig:PVgradients} are both suggestive of a role
for barotropic instability during spinup of the superrotating state. \figref{fig:barotropic} displays a
schematic of the mechanism of barotropic instability for an idealized zonal
mean zonal wind following the initial, nearly axisymmetric phase of the
simulation. Rossby waves form in regions of vorticity gradients and if the
vorticity gradient is strong these are essentially edge waves. If a pair of
edge waves form at the equator and at higher latitudes and have sufficient
overlap in space, they are able to phase lock and interact to form an
instability. This gives rise to the circulation depicted in
\figref{fig:barotropic} by slanted ellipses, with the instability `leaning into
the shear' in order to extract energy from it. Since the zonal wind speed is
higher at high latitudes than at low latitudes, phase locking of the waves
requires them to travel in opposite directions relative to the local mean flow,
and for Rossby waves this would require their local PV gradients ($\ppp qy$) to be of
opposite sign. Because the circulation formed by the instability slants into
the shear the Reynolds stresses from these disturbances transfer eastward
momentum down the wind shear, converging momentum at the equator and producing
superrotation. Furthermore, because the interacting edge waves must span a
large meridional distance, the zonal scale of the waves is correspondingly
large, so explaining the emergence of wavenumber one in our simulations.

The above picture is, however, an oversimplification, even if it captures the basic underlying mechanism leading to momentum convergence at the equator. At the onset of the instability (for example, at day 220 in \figref{fig:PVsnapshots}), the potential vorticity gradient is positive at the equator and becoming negative in the subtropics before changing sign again to be large and positive at high latitudes. These large scale reversals in potential vorticity gradient with latitude are certainly indicative of a barotropic instability, although it is not only the horizontal shear and the beta term that contribute to the potential vorticity; rather, terms involving the stratification and vertical shear are equally important (not shown).

Given the large-scale nature of the instability, the reason for the disruption of the development of superrotation by a sponge layer placed at a particular latitude band becomes clear (see \figref{fig:spongehadleywinds}).  The structure of the global disturbance is significantly altered if any portion of it is damped, thus compromising the wave's ability to create momentum flux divergence.  We found the disruption to be particularly effective for a thin sponge layer placed at the equator, consistent with an equatorial role in the instability giving rise to superrotation.

\begin{figure*}[tb]
\begin{center}
\includegraphics[width=5in]{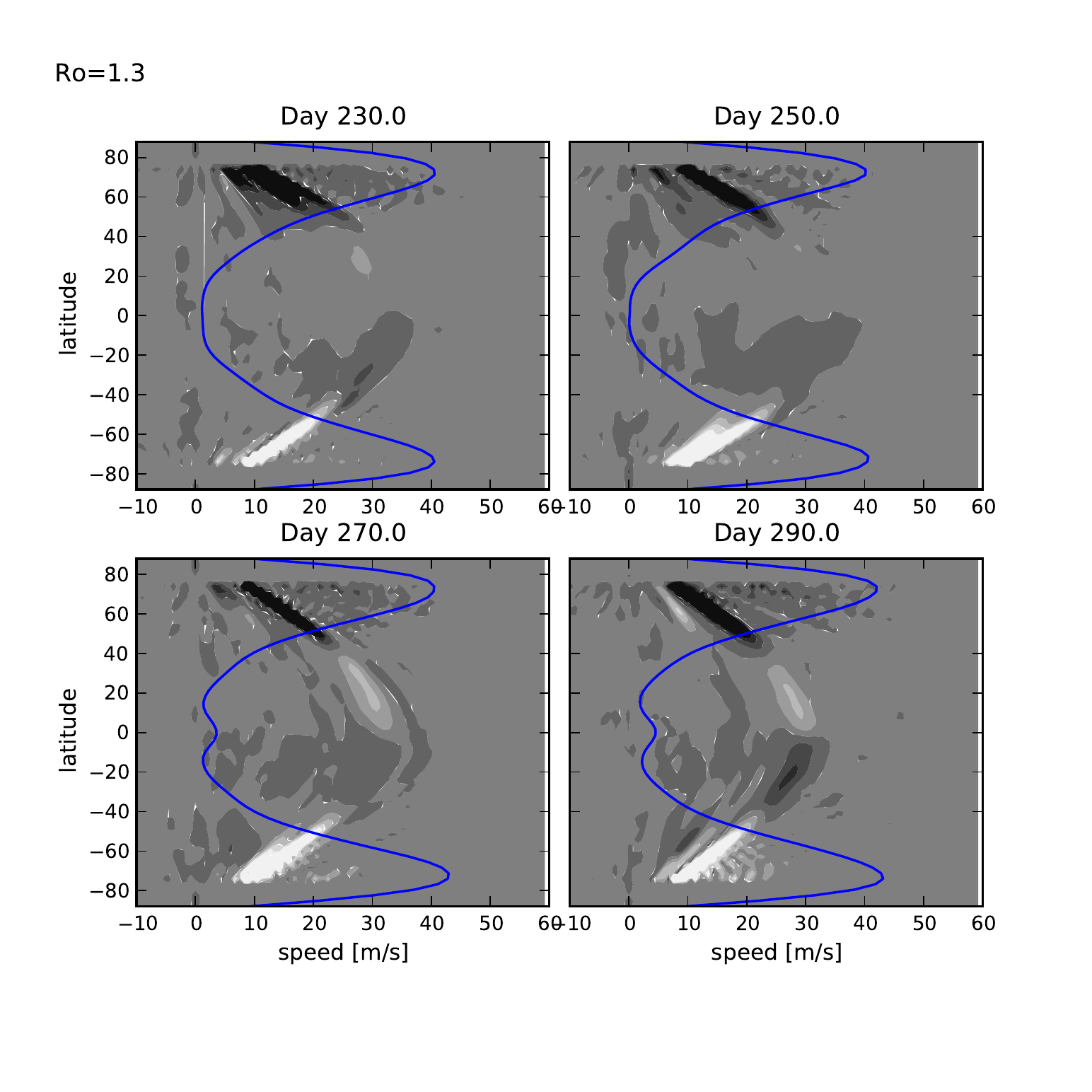}
\caption{Same as \figref{fig:Ro10phasespeedSpectra} for the \Rot= 1.3 case during spinup.}
\label{fig:Ro1phasespeedSpectra}
\end{center}
\end{figure*}

We can also see what prevents strong superrotation from developing in the $\Rot=  1.3$ case.  \figref{fig:Ro1phasespeedSpectra} displays a sequence of phase-speed--latitude cospectra (shaded) with the zonal mean zonal wind overplotted (line) for the \Rot=  1.3 case.  At high latitudes, several Rossby waves that propagate westward relative to the mean flow are evident.  These disturbances are due to baroclinic instability, and they converge westward (retrograde) angular momentum to critical layers throughout the entire latitudinal domain.  A global wave transiently appears and accelerates the equatorial winds for a time, but the restoring torque of the high-latitude waves breaking at lower latitudes dominates the low-latitude momentum budget, keeping equatorial winds weakly prograde.

\section{Superrotation in the Solar System} \label{sec:solarsystem}

The atmospheres of two of the four terrestrial bodies in the Solar System with thick atmospheres --- Titan and Venus --- are observed to superrotate, whereas the atmospheres of Earth and Mars do not.  We have shown the thermal Rossby number is a control parameter in the transition to superrotation, and this parameter is most sensitive to rotation rate and radius.  Titan and Venus respectively have rotation periods 16 and 243 times that of Earth; Venus is nearly Earth's twin in size, while Titan's radius is $\sim$40\% Earth's,
and consequently the thermal Rossby numbers of Titan and Venus are both much greater than one.   Since the transition to superrotation occurs for values of \Rot exceeding one, Titan and Venus lie well beyond this transition in parameter space. It is possible that Venus (with $\Rot \approx 1200$) is in yet another parameter regime from that explored in this paper, but the essential dynamics of the superrotation of Titan (with $\Rot \approx 30$) may be similar to those occurring in our simulations (this connection, and the role of seasonality, will be further explored in a subsequent paper). The radius of Mars is roughly 50\% of Earths and has a nearly identical rotation rate.  Therefore, $\Rot(\text{Mars}) \sim 0.1$, and so Mars is the closest to being at the superrotation transition; a further reduction in the Martian rotation rate could cause the atmosphere to superrotate.  However given Mars's lack of a large satellite and weak tidal coupling with the Sun, it is unlikely that the rotation rate has changed significantly over the lifetime of the Solar System (or will change in the foreseeable future). Superrotation on the gas giants seems likely via mechanisms that do not involve a global mode but that may involve convection at the equator \citep{Yamazaki_etal05,Lian_Showman09,Schneider_Liu09} or deep convection \citep{Heimpel_Aurnou07,Kaspi_etal09}.

\section{Conclusions} \label{sec:conc}

We have described the mechanism of transition of a terrestrial atmosphere to
superrotation that occurs when \Rot, the thermal Rossby number, exceeds unity, as is
characteristic of the atmospheres of Venus and Titan. The Earth's equatorial
atmosphere (with $\Rot \ll 1$) does not superrotate because there is no
mechanism to excite the large-scale disturbance we have identified in our \Rot=
10.5 case. Rather, Rossby waves that are small compared to the domain size are formed at mid-latitudes, propagate
equatorward and break, decelerating the equatorial flow and damping any
tendency toward superrotation that might be present. However, if \Rot exceeds
unity a robust mechanism emerges: the Rossby wave source in midlatitudes
weakens (because the scale of baroclinic instability becomes larger
than the planetary circumference) and a global disturbance predominantly of
zonal wavenumber one develops that provides a convergence of momentum at the
equator.  Both baroclinic and barotropic instability play a role in the spinup phase, while barotropic instability alone is responsible for the maintenance of superrotation in the statistically steady-state.

The global disturbance during spinup of our \Rot = 10.5 case as the first baroclinic mode is associated with high-latitude PV gradient reversal in the vertical direction (suggestive of baroclinic instability) as well has PV gradient reversals in the horizontal direction, near the equator (suggestive of barotropic instability).   The disturbance deposits eastward (prograde) momentum in the equatorial region, accelerating the zonal wind into a superrotating state. The PV gradient changes sign in the horizontal direction near the equator, indicating a role for barotropic instability.  Disturbances associated with these near-equatorial regions overlap, giving the equatorial portion of the mode the appearance of a chevron that is symmetric about the equator.  The spinup mechanism thus involves a mixed baroclinic-barotropic instability; the relative importance of these needs further study.

Superrotation is maintained in the steady-state by a persistent, global barotropic mode associated with a mixture of high- and low-latitude regions of PV gradient reversal in the horizontal (indicating barotropic instability).  Equatorial eddy momentum flux convergence by the global barotropic mode is generally disorganized and weak.  However, frictional and other torques tending to decelerate mean zonal winds at the equator and above the boundary layer are also quite weak, so little is required of the global disturbance in order to maintain the superrotating state. 

The heuristic picture of linear wave propagation and critical layer absorption is in contrast with the global coherence of the dominant waveforms in our high-Rossby-number simulations.  The disturbances responsible for the transition to superrotation are global in scale and thus are unable to propagate.  Rather than providing a critical layer for a propagating wave, the zonal mean latitudinal shear plays an important role in the meridional transport of momentum by its interaction with the global wave.   

The dominance of long-wavelength disturbances in our results suggest modeling studies of Titan aimed at developing realistic levels of superrotation should not require high resolution.  On the other hand, we demonstrated the development of superrotation in our model is strongly sensitive to damping at the equator.  Models having unrealistic dissipation at or near the equator will thus have difficulty producing and maintaining superrotation.  

For intermediate \Rot, there is a competition between equatorial acceleration by
global waves and deceleration by Rossby waves due to mid- and high-latitude
baroclinic instability. Our \Rot= 1.3 case shows evidence for transient global
waves like that seen in the \Rot= 10.5 case, but also displays baroclinic instability at mid-latitudes. Strong superrotation does not
develop because of the deceleration caused by the breaking of meridionally propagating Rossby waves generated at mid- and high-latitudes by baroclinic instability. 
 
Equatorial waves with a characteristic chevron shape similar to the equatorial portion of the global wave in our simulations are observed in Venus' cloud layer, and these waves may be connected with the mechanism described here.  If the observed waves correspond to the equatorial component of the global wave we have described, then a high-latitude component should also be present whose geopotential anomaly varies coherently with respect to the equatorial component (see \figref{fig:globalwave2}).  Similar waves have not yet been observed in Titan's atmosphere, but may nonetheless be observable from time to time in thermal infrared data from the Cassini orbiter.

Of course, notwithstanding the heuristic description that we have given, the mechanism we have identified remains to be analytically described in terms of a classical linear instability or wave analysis that would yield its spatial structure and dispersion relation in terms of the fundamental parameters of the flow, in a manner akin to the classical analyses of the eponymous waves and instabilities of Rossby, Charney and Eady.  

\bibliographystyle{agu04}

\end{article}

\end{document}